\documentclass[aps,prx,twocolumn,superscriptaddress,showpacs,floatfix,longbibliography]{revtex4-1}
\usepackage{amsmath,amssymb,graphicx}

\usepackage[utf8]{inputenc}
\usepackage[T1]{fontenc}
\usepackage{xcolor}
\usepackage{dsfont}
\usepackage{amsthm,amsmath,amssymb}
\usepackage{mathrsfs}

\IfFileExists{newtxtext.sty}
{\usepackage{newtxtext,newtxmath}}
{\IfFileExists{stix.sty}
	{\usepackage{stix}}
	{\IfFileExists{mathptmx.sty}
		{\usepackage{mathptmx}}{} } }

\usepackage{textcomp}

\usepackage{bm}

\IfFileExists{siunitx.sty}{\usepackage{booktabs,siunitx}}{}

\pdfoutput=1
\usepackage{color}
\definecolor{LinkColor}{rgb}{0.256,0.439,0.588}
\usepackage{hyperref}
\hypersetup{
	pdfauthor={Wei Wang},
	pdftitle={paper},
	colorlinks=true,
	citecolor=LinkColor,
	linkcolor=LinkColor,
	urlcolor=LinkColor
}

\newcommand{\beq} {\begin{equation}}
\newcommand{\eeq} {\end{equation}}
\newcommand{\bea} {\begin{eqnarray}}
\newcommand{\eea} {\end{eqnarray}}
\newcommand{\be} {\begin{equation}}
\newcommand{\ee} {\end{equation}}
\renewcommand{\(}{\left(}
\renewcommand{\)}{\right)}

\begin{document}
\title{Phase diagram of the spin-1/2 Yukawa-SYK model: Non-Fermi liquid, insulator, and superconductor}
\author{Wei Wang}\thanks{These two authors contributed equally.}
\affiliation{Beijing National Laboratory for Condensed Matter Physics and Institute of Physics, Chinese Academy of Sciences, Beijing 100190, China}
\affiliation{School of Physical Sciences, University of Chinese Academy of Sciences, Beijing 100190, China}
\author{Andrew Davis}\thanks{These two authors contributed equally.}
\affiliation{Department of Physics, University of Florida, Gainesville, FL 32601}
\author{Gaopei Pan}
\affiliation{Beijing National Laboratory for Condensed Matter Physics and Institute of Physics, Chinese Academy of Sciences, Beijing 100190, China}
\affiliation{School of Physical Sciences, University of Chinese Academy of Sciences, Beijing 100190, China}
\author{Yuxuan Wang}
\email[]{yuxuan.wang@ufl.edu}
\affiliation{Department of Physics, University of Florida, Gainesville, FL 32601}
\author{Zi Yang Meng}
\email[]{zymeng@hku.hk}
\affiliation{Department of Physics and HKU-UCAS Joint Institute of Theoretical and Computational Physics, The University of Hong Kong, Pokfulam Road, Hong Kong SAR, China}
\affiliation{Beijing National Laboratory for Condensed Matter Physics and Institute of Physics, Chinese Academy of Sciences, Beijing 100190, China}

\begin{abstract}
We analyze the phase diagram of the spin-1/2 Yukawa-Sachdev-Ye-Kitaev model, which describes complex spin-1/2 fermions randomly interacting with real bosons via a Yukawa coupling, at finite temperatures and varying fermion density. In a recent work ~[PhysRevResearch.2.033084] it has been shown that, upon varying filling or chemical potential, there exists a first-order quantum phase transition between a non-Fermi liquid (nFL) phase and an insulating phase. Here we show that in such a model with time-reversal symmetry this quantum phase transition is preempted by a pairing phase that develops as a low-temperature instability. As a remnant of the would-be nFL-insulator transition, the superconducting critical temperature rapidly decreases beyond a certain chemical potential. On the other hand, depending on parameters, the first-order quantum phase transition extends to finite-temperatures and terminate at a thermal critical point, beyond which the nFL and the insulator become the same phase, similar to that of the liquid-gas and metal-insulator transition in real materials. We determine the pairing phase boundary and the location of the thermal critical point via combined analytic and quantum Monte Carlo numeric efforts. Our results provide the model realization of the  transition of nFL's towards superconductivity and insulating states, therefore offer a controlled platform for future investigations of the generic phase diagram that hosts nFL, insulator and superconductor and their phase transitions.
\end{abstract}

\maketitle

\section{Introduction}
To understand the non-Fermi liquid (nFL) behavior of interacting electron systems is one of the central issues in modern condensed matter physics. Widely believed to be relevant to the microscopic origin of the strange metal behavior in unconventional superconductors~\cite{Keimer2015, ZhaoyuLiu2016, YanhongGu2017,Custers2003,BinShen2019,YCao2019,shen2019observation,ChuangChen2020}, its theoretical description remains a challenging issue due to the lack of a small control parameter. Recently, the Sachdev-Ye-Kitaev (SYK) model has garnered widespread attention as it emerges as a new paradigm for the study of nFLs~\cite{SY,K,SYK2,SYK3}, which is different from most previous research of nFLs, where the system is usually realized in itinerant
fermions coupled to soft bosonic modes near a quantum critical point~\cite{abanov2003,Metlitski2010a,Metlitski2010b,PhysRevB.98.045116,Liu2019PNAS,XiaoYanXuReview2019,XiaoYanXu2020,Damia2020}. The nFL in SYK model is exactly solvable in the large-$N$ limit. Beyond the context of non-Fermi liquids, the SYK model also has been found to have a hidden
holographic connection to quantum black holes and saturates the
limiting rate of scrambling due to its short equilibration time~\cite{guo-gu-sachdev-2020,YingfeiGu2017}.

Motivated by the aforementioned fermionic systems near quantum-critical points, recently, a variant of the SYK model, dubbed the Yukawa-SYK model, has been proposed~\cite{PhysRevLett.124.017002,PhysRevB.100.115132,HAUCK2020168120,2020Yukawa,PhysRevResearch.2.033084}. Such a model describes strong random Yukawa coupling between $MN$ flavors of fermions and $N^2$ flavors of bosons. Analytical investigation at large-$N$ has revealed a saddle point solution in which the Yukawa coupling ``self-tunes" the massive bosons to criticality and the fermions form a nFL, which saturates the bound on quantum chaos~\cite{Kim-Cao-Altman,Kim-Alterman-Cao}. This saddle point solution has been verified at finite $N$ and finite $T$ via quantum Monte Carlo (QMC) simulations with an additional antiunitary time-reversal symmetry~\cite{2020Yukawa} by making use of a reparametrization symmetry of the large-$N$ solution.

In this work we focus on the fate of the nFL upon varying temperature and density. Similar to the complex SYK model~\cite{PhysRevLett.120.061602,smit2020quantum}, it has been recently shown that there exists a first-order quantum phase transition at finite chemical potential separating a nFL state and a gapped insulating state~\cite{PhysRevResearch.2.033084}. On the other hand, in several versions of the Yukawa-SYK model and the complex SYK model, the nFL phase becomes unstable to a pairing phase~\cite{PhysRevB.100.115132,HAUCK2020168120,PhysRevResearch.2.033025,PhysRevLett.124.017002,PhysRevB.101.184506,PhysRevB.100.220506,PhysRevB.100.115132,HAUCK2020168120,PhysRevLett.124.017002}. In particular for the Yukawa-SYK models~\cite{PhysRevLett.124.017002,PhysRevB.100.115132,HAUCK2020168120,2020Yukawa,PhysRevResearch.2.033084}, two distinct pairing behaviors have been reported, depending on whether fermions carry spin degrees of freedom. For the spinless Yukawa-SYK model studied in Ref.~\cite{PhysRevLett.124.017002}, it was found that pairing only occurs for a certain range of the ratio between boson and fermion flavor numbers, while for spin-1/2 Yukawa-SYK models~\cite{PhysRevB.100.115132,HAUCK2020168120,2020Yukawa}, the nFL state is in general unstable toward pairing at sufficiently low temperatures.

\begin{figure}[!htp]
\centering
\includegraphics[width=0.9\columnwidth]{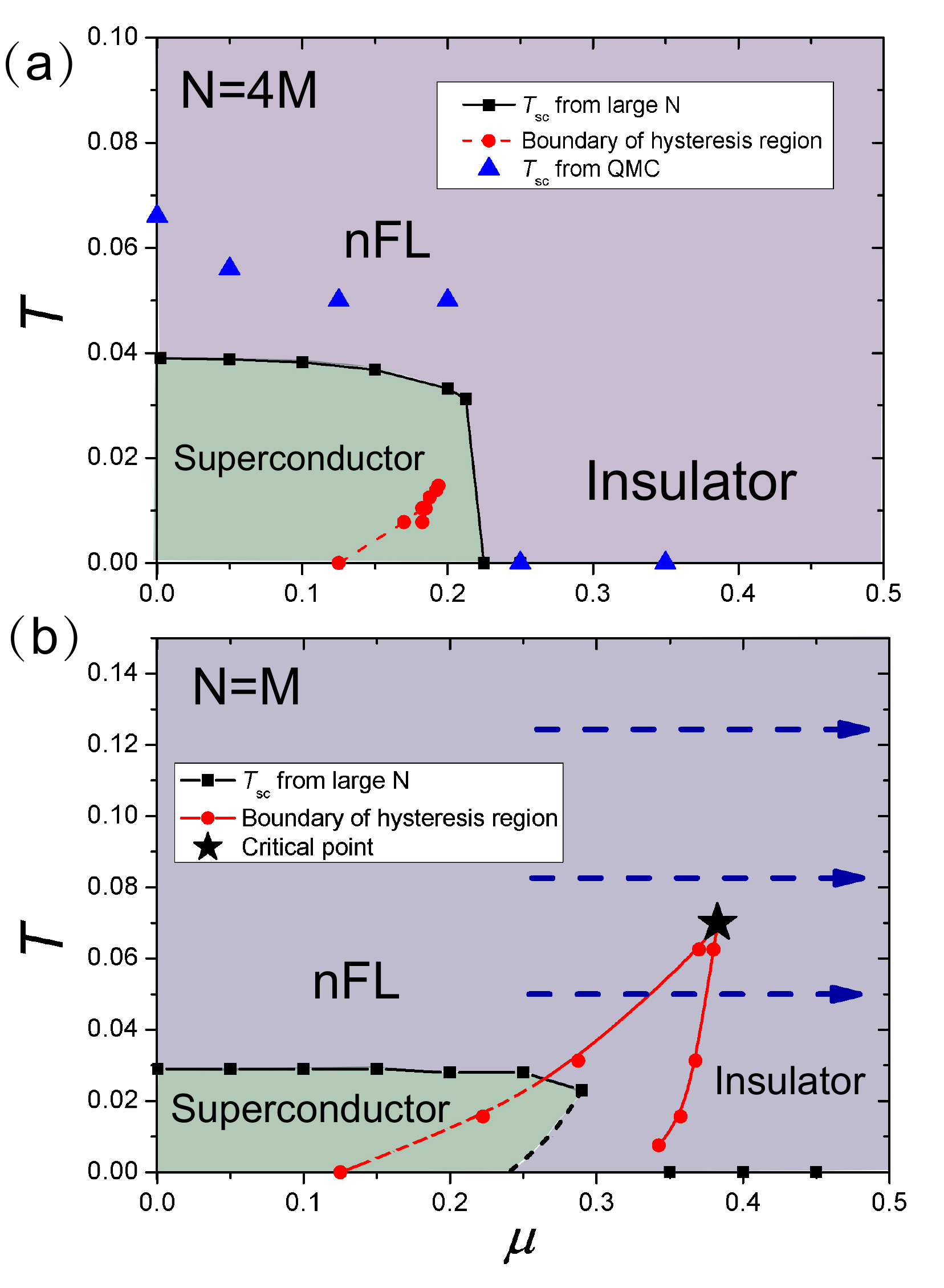}
\caption{(a) Phase diagram of the Yukawa-SYK Model at $N=4M,\omega_0=1,m_0=2$. From the large-$N$ calculation, one sees the nFL become superconductor at low temperature in a wide range of chemical potential, and the first-order hysteresis region denoted by the red points and lines, obtained in the absence of pairing instability within large-$N$. In the presence of superconductivity the position of this region is renormalized; hence the dashed line. The thermal critical point that terminates the first order transition locates at $(\mu_c=0.194,T_c=0.015)$. The blue triangles are the transition points from nFL to superconductor obtained from QMC at finite $N,M$ (cf. Fig.~\ref{fig:fig4}), which are consistent with the results obtained from large-$N$ calculations (black squares). 
(b) Phase diagram of the Yukawa-SYK Model at $N=M,\omega_0=1,m_0=2$ from the large-$N$ calculation. The first-order hysteresis region denoted by the red points and lines are obtained in the absence of pairing instability within large-$N$. The dashed-line portion of this boundary is renormalized by superconductivity phase. The black dashed line denotes the boundary of the superconducting phase within the hysterisis region and is only depicted qualitatively.  The QMC $n-\mu$ curves in Fig.~\ref{fig:fig2} are along blue dashed paths. The thermal critical point at $(\mu_c=0.3825,T_c=0.07)$ is denoted by the black star.}
\label{fig:fig1}
\end{figure}

Here, by means of large-$N$ calculation and unbiased large scale QMC simulation, we present the global phase diagram (see Fig.~\ref{fig:fig1}) of the spin-1/2 version of the Yukawa-SYK model, spanned by the axes of  temperature $T$ and chemical potential $\mu$. Up to a critical value in the chemical potential $\mu$,  a finite temperature phase transition from nFL to  superconductivity is observed. We determine the pairing transition from solving the linear Eliashberg equation using large-$N$ result of the Green's functions, as well as finite-size scaling of the pairing susceptibility in QMC simulations. We obtain a good agreement between the two methods, indicating the pairing transition is mean-field like.
In particular, in the weak coupling limit, we analytically determine the threshold value for $\mu$ for the superconductor-insulator transition at zero temperature, which agrees well with numerical results. On the other hand, by solving the Schwinger-Dyson equation, we found the first-order quantum phase transition extends to low temperature and terminates at a (thermal) critical point, which is a generic feature in many metal-insulator transition in correlated materials~\cite{Imada1998,Limelette2003}. However, depending on the strength of the first-order quantum transition (previously found to be controlled by the ratio $M/N$~\cite{PhysRevResearch.2.033084}), we show that the thermal critical point may be masked by the superconducting phase.  The phase diagram obtained offers a controlled platform for future investigations of phase transitions between nFL, insulator and superconductor, at generic electron fillings.

\section{The spin-1/2 Yukawa-SYK model}
The Yukawa-SYK Model we study is described by the following Hamiltonian,
\begin{eqnarray}
H &=& \sum_{i,j=1}^{M}\sum_{\alpha,\beta=1}^{N} \sum_{m,n=\uparrow,\downarrow}(\frac{i}{\sqrt{MN}} t_{i\alpha,j\beta}\phi_{\alpha\beta}c^\dagger_{i\alpha  m}\sigma_{m,n}^z c_{j\beta n}) \nonumber\\
&+&\sum_{\alpha , \beta =1}^N(\frac{1}{2}\pi_{\alpha\beta}^2+\frac{m_0^2}{2}\phi_{\alpha\beta}^2)
-\mu\sum_{i=1}^M \sum_{\alpha=1}^N\sum_{m=\uparrow,\downarrow} c_{i\alpha m}^\dagger c_{i\alpha m},
\label{eq:eq1}
\end{eqnarray}
where $c_{i\alpha m}$($c_{i\alpha m}^\dagger$) is the annihilation (creation) operator for a fermion with flavor $\alpha$ and spin $m,n$ ($\uparrow$ or $\downarrow$). 
The random Yukawa coupling parameter between fermion and boson is realized as $\langle t_{i\alpha,j\beta}\rangle=0$, $\left\langle t_{i \alpha, j \beta} t_{k \gamma, l \delta}\right\rangle=\left(\delta_{\alpha \gamma} \delta_{i k} \delta_{\beta \delta} \delta_{j l}+\delta_{\alpha \delta} \delta_{i l} \delta_{\beta \gamma} \delta_{j k}\right) \omega_{0}^{3}$. 
We set $\omega_0=1$ as the energy unit throughout the paper. The dynamical behavior of the boson has been given in the second term and $\pi_{\alpha \beta}$ is the canonical momentum of $\phi_{\alpha \beta}$. 
Hermiticity of the model requires $\phi_{\alpha \beta}=-\phi_{\beta \alpha}$. $(\alpha, \beta)$ are flavor indices which run from 1 to $N$ and $(i,j)$ are quantum dot indices which run from 1 to $M$. $\sigma^z$ represents the $z$ component of the fermion spin. Due to time-reversal symmetry, this Hamiltonian is free from the fermion sign problem and can be simulated by QMC at finite $M$ and $N$ and at finite doping with $\mu\neq0$. We prove the absence of the sign problem and discuss the QMC implementation in Appendix A.

As we mentioned, compared to the spinless Yukawa-SYK model previously studied~\cite{PhysRevLett.124.017002} the key difference is the inclusion of spin degree of freedom,
which enables sign-problem-free quantum Monte Carlo simulation of the model. Physically, this modification
introduces an instability toward spin-singlet pairing, while in
Ref.~\cite{PhysRevLett.124.017002} the pairing of spinless fermions only occurs at certain
regimes of $(M,N)$. The behavior of the model \eqref{eq:eq1} at $\mu=0$ was studied in our previous work in Ref.~\cite{PhysRevResearch.2.033084}, and in this work we focus on the phases for a generic $\mu$.

The main results of this work can be summarized by the two representative phase diagrams. We found that in general there exist a superconducting dome $T_{sc}(\mu)$ in $T-\mu$ plane shown in Fig.~\ref{fig:fig1} (the phases for positive and negative $\mu$ are identical by particle-hole symmetry). The vanishing of pairing for larger $\mu$ is driven by the underlying nFL/insulator transition. In the absence of pairing, we obtain a hysteresis region (the wedge region marked by red lines in Fig.~\ref{fig:fig1}) in which both nFL and insulator states are metastable, divided a first-order phase transition inside the wedge, similar to that of the liquid-gas transition and metal-insulator transition in many correlated materials~\cite{Imada1998,Limelette2003}.  The exact location of the first-order transition requires comparing the free energy for different solutions, which is beyond the scope of the current work.
For $N=4M,\omega_0=1,m_0=2$, the superconducting dome completely preempts the would-be nFL/insulator transition, while for $N=M,\omega_0=1,m_0=2$, the first-order phase transition is stronger and the corresponding thermal critical point occurs outside of the superconducting phase. In Fig.~\ref{fig:fig1}(a) we have also marked the superconducting critical temperature obtained by finite-size scaling from QMC data with blue triangles, which are consistent with the results obtained from large-$N$ calculations denoted by the black squares. The exact position of the hysteresis region inside the superconducting phase, and vise versa, requires solving the non-linear superconducting gap equation, and is qualitatively marked by dashed lines respectively.
\begin{figure}[!htp]
\centering
\includegraphics[width=\columnwidth]{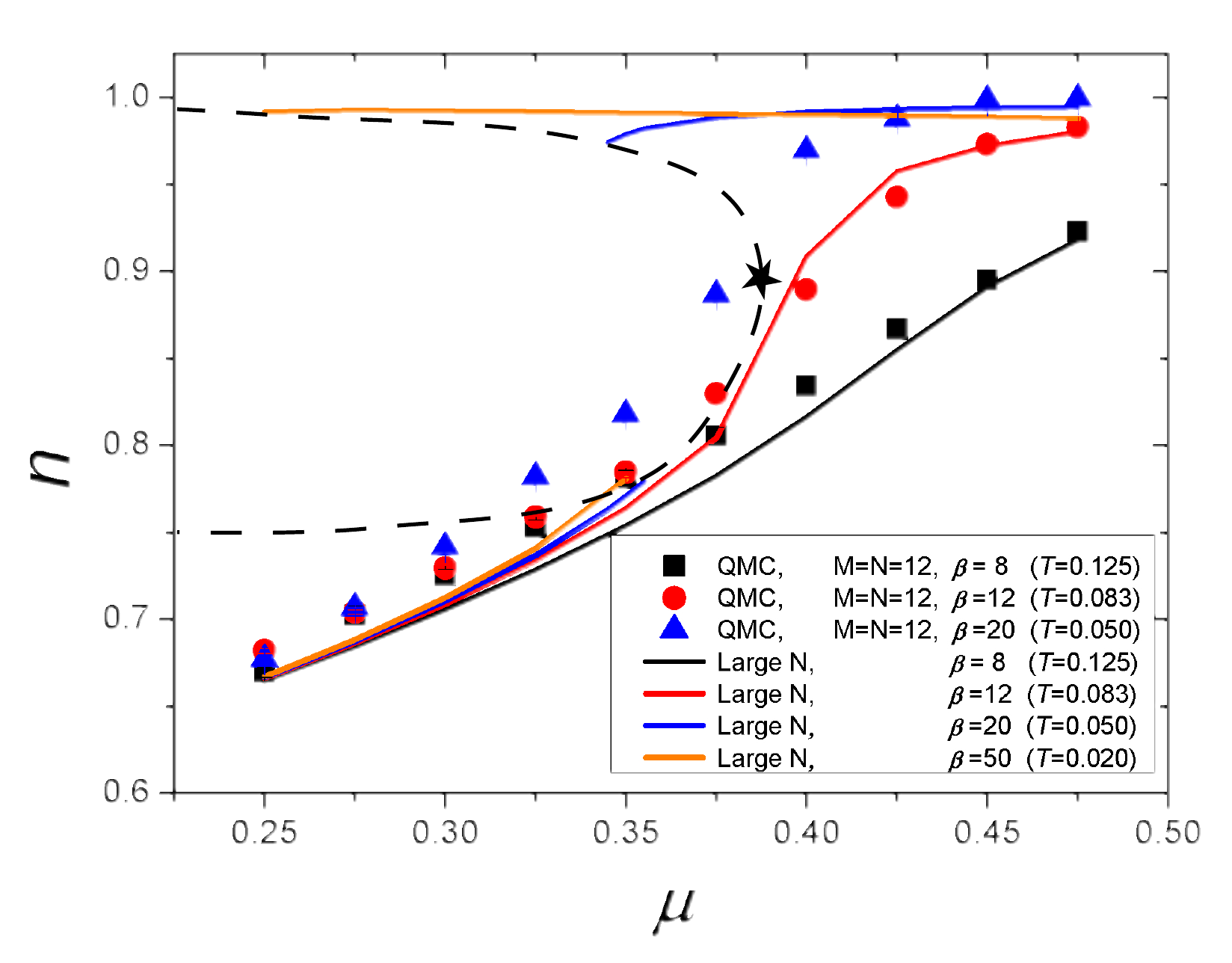}
\caption{Filling $n$ versus $\mu$ for selected $T$ in the vicinity of the thermal critical point in Fig.~\ref{fig:fig1} (b) for $N=M, \omega_0=1, m_0=2$. At higher $T=0.125$, one sees the $n(\mu)$ curves both from large-$N$ and QMC are smooth. $T=0.125, \mu=0.375$ (marked by the star), is a thermal critical point. At temperature $T=0.083$ close to the thermal critical point, there is a sharp turn in of $n(\mu)$ signifying the divergence of the compressibility $dn/d\mu$. At lower $T$, there is a range of $\mu$---the hysteresis region---where the filling is double-valued. The lower branch represents the nFL behavior and the upper branch represents the insulating behavior; a first order transition connects the two at a chemical potential given by a Maxwell construction. The dashed line delimits the region where solutions are unstable. Note that the QMC results at $T=0.050$ (the blue triangles) further differ from
the first order transition behavior (the blue curves). At finite $M,N$ and at finite temperatures, the system is finite and does not have phase transitions.
Instead the system undergoes crossovers, which become phase transitions only in the thermal dynamical
limit. We have verified that numerically as one increase $M,N$ the numerical data indeed tends to approach the large-$N$ curve. }
\label{fig:fig2}
\end{figure}

\section{ Phases in the normal state}
 Within large-$N$, we  map out the $T$-$\mu$ phase diagram by numerically solving the Schwinger-Dyson Eqs.~\eqref{eq:eq2}. In terms of the propagators $G_f^{-1}(i\omega) = i\omega + \mu + \Sigma(i\omega)$ for the fermions and $G_b^{-1}(i\Omega) = \Omega^2 + m_0^2 + \Pi(i\Omega)$ for the bosons, the Schwinger-Dyson equations are
\begin{eqnarray}
&\Sigma(i\omega) = -\omega_0^3 &\int \frac{d\Omega}{2\pi} G_b(i\Omega)G_f(i\omega-i\Omega), \nonumber\\
&\Pi(i\Omega) = \frac{4M}{N} \omega_0^3 &\int \frac{d\omega}{2\pi} G_f(i\Omega/2+i\omega) G_f(-i\Omega/2+i\omega).
\label{eq:eq2}
\end{eqnarray}
Since we work in the large-$M,N$ regime, and only the ratio $M/N$ enters the equations~\cite{2020Yukawa}.

We solve Eqs.~\eqref{eq:eq2} iteratively, by starting with a simple ansatz for $\Sigma_0$ and $\Pi_0$ on the right-hand-sides, obtaining updated values $\Sigma_1$ and $\Pi_1$ on the right-hand-sides, and repeating until the solutions $\{\Sigma_{n}\}$ and $\{\Pi_{n}\}$ saturate, where $n$ is the iteration step number. Noting that Eqs.~\eqref{eq:eq2} is consistent with the assumptions that $\Pi(i\Omega)$ is even and that the real and imaginary parts of $\Sigma(i\omega)$ are even and odd, respectively, we need only compute the self-energies at nonnegative frequencies. However, directly implementing this strategy leads to divergent behavior, especially at $\Sigma_n(\pm \pi T)$ and $\Pi_n(0)$. This issue is related to the fact that $\Sigma_n(\pm \pi T)$ and $\Pi_n(0)$ are determined by the behaviors of $G$ and $D$ at all frequency scales, rather than their ``local" behavior at nearby low energies~\cite{PhysRevResearch.2.033084}. Indeed, in analytical solutions of Eq.~\eqref{eq:eq1} at $T=0$~\cite{PhysRevResearch.2.033084}, the conditions on $\Sigma(0)$ and $\Pi(0)$ were used to determine the ultraviolet energy scale beyond which nFL behavior crosses over to that of a free system.  To avoid the instability at lowest frequency points in the iterative method, we artificially introduce the ``stabilizers" for each step of the iteration by uniformly shifting $\Sigma_{n}$ and $\Pi_n$ such that
\begin{align}
\Sigma'_{n}(\pi T) = s,~~\Pi'_n(0)=p.
\end{align}
This prescription prevents $\Sigma_{n}(\pi T)$ and $\Pi_n(0)$ from running away.
Of course, in general after the iteration converges, the solution we get is \emph{not} the solution of the original SD equation, unless the updated value  at the next step coincides with the stabilizers, i.e., $\Sigma_{n+1}(\pi T)\to s$ and $\Pi_{n+1}(0)\to p$. Using this criterion, we can find the correct values of the stabilizers $s_0$ and $p_0$. The necessary shifts are typically extremely small compared to the values of the self-energies over the frequency range where most of their support lies.

At low $T$ and for some ranges of $\mu$, we obtain two different choices of stabilizers $\{s_0,p_0\}$ which cause the iteration to converge. This signals the hysteresis behavior, and the resulting two types of solutions physically correspond to nFL and insulator behaviors that are local minimum of the free energy. This method does not reproduce the unstable ($dn/d\mu < 0$) solutions, whose boundary  are sketched  qualitatively in Fig. ~\ref{fig:fig2}. The filling $n$ is calculated from the imaginary-time fermionic Green's function
\begin{equation}
n = \frac{1}{2} + \sum_{\omega_n} G_f(i\omega_n)
\end{equation}
Alternatively, one can obtain the filling from $n=\lim_{\delta\to 0}G(\tau=-\delta)$. However, due to the finite number of frequency points we keep, $G(\tau)$ obtained from a Fourier transform exhibits strong oscillations at small $\tau$ (the Gibbs phenomenon).

From the fermion Green's functions, we extract the $n-\mu$ curve, shown in Fig.~\ref{fig:fig2}. In particular, for certain values of $(\mu, T)$ the two solutions coexist, indicating the existence of metastable states. At low temperatures, there is in general a range of $n$ for which no solutions were found. Nevertheless, we expect in the full solution the $n(\mu)$ the curve to be {smooth}. This missing portion of solutions (see the dashed lines in Fig.~\ref{fig:fig2}) thus correspond to those that cannot be obtained from a stable convergent iterative series. We thus identify this missing portion as thermodynamically unstable saddle points of the free energy. Such a behavior is typical of first-order phase transitions. Like water-vapor transition, the actual $n-\mu$ curve connecting the two branches is a straight line determined by the Maxwell construction. Above a certain temperature $T_c$, the two types of solutions become smoothly connected at a chemical potential $\mu_c$.  Here the compressibility $dn/d\mu$ diverges, and thus $(T_c,\mu_c)$ is a thermal critical point of the system.

{Qualitatively, the value of $T_c$ up to which the first-order phase transition survives is related to the strength of \emph{quantum} first-order transition at $T=0$. In Ref.~\cite{PhysRevResearch.2.033084}, it was obtained analytically that the first-order quantum phase transition is weaker for a larger ratio $M/N$ and becomes continuous at $M/N\to \infty$. Indeed, we obtain that $T_c$ for the case $N=M$ is higher than that with $N=4M$. }

Here we note that the normal state phases of the spinless Yukawa-SYK model~\cite{PhysRevLett.124.017002} can be obtained by a similar analysis. Indeed the only difference is an additonal factor of two  in the second equation of \eqref{eq:eq2} coming from summing over spin species. However, as we see below, the pairing phase of the spin-1/2 Yukawa-SYK model comes from the spin-singlet channel, which is absent in the spinless version, as was discussed in Ref.~\cite{PhysRevLett.124.017002}.

%
%

\section{Pairing transition}
{The interaction mediated by the boson exchange is attractive in the equal-index, spin-singlet Cooper channel~\cite{2020Yukawa}, and the system has an instability a low-temperature pairing phase.}
The Eliashberg equation is given by
\begin{align}
\Phi(\omega_n) = &\omega_0^3 T \sum_{\Omega_m} G_b(i\Omega_m) G_f(i(\omega_n + \Omega_m)) G_f(-i(\omega_n + \Omega_m))\nonumber\\
&\times \Phi(\omega_n+\Omega_m).
\label{eq:eq3}
\end{align}

{At $\mu=0$, the pairing problem has been analyzed in Ref.~\cite{2020Yukawa}. For $\mu\neq 0$, due to the breaking of particle-hole symmetry, the mismatch between $G(\pm i\omega_n)$ leads to a reduced pairing tendency, much like a Zeeman splitting  in momentum space reduces the spin-singlet pairing susceptibility. We can glean some insight about the pairing transition by considering the weak-coupling limit $\omega_0\ll m_0$ and determine the value of $\mu_{sc}$ beyond which pairing vanishes.
At $T=0$}, the Schwinger-Dyson equations admit an insulating solution approximated~\cite{PhysRevResearch.2.033084} by $\Sigma(i\omega) = -\omega_F/2$, where $\omega_F\equiv \omega_0^3/m_0^2$, and $\Pi(i\Omega) = 0$ as long as $\mu > \omega_F/2$. These self-energies become exact in the limit. In this regime the pairing equation becomes
\begin{equation}
\Phi(\omega) = \omega_0^3 \int \frac{d\Omega}{2\pi}
\frac{1}{(\Omega-\omega)^2 + m_0^2} \frac{1}{\Omega^2 + (\mu-\omega_F/2)^2} \Phi(\Omega)
\end{equation}
Most of the support of the integral comes from frequencies on the order of $\omega_F$, so at very weak coupling the frequency dependence of the boson propagator can be ignored. With the ansatz $\Phi(\omega) = \textrm{const}$, performing the integral reveals a pairing transition at
\be
\mu_{sc} = \omega_F \(\equiv \omega_0^3/m_0^2\).
\label{eq:musc}
\ee
To verify this analytic result, we solved the pairing equation numerically at very weak coupling: $m_0 = 10 \omega_0$, and we obtained $\mu_{sc} \approx 0.98 \omega_F$ for the maximum chemical potential beyond which pairing vanishes, and we also found that $\Phi(\omega)$ is virtually constant, justifying our ansatz. Extrapolating to our case with $\omega_0/m_0=0.5$, we expect $\mu_{sc}=0.25$. This indeed matches well with the numerical results from large-$N$ (Fig.~\ref{fig:fig1}(a,b)) and from QMC (Fig.~\ref{fig:fig1}(a)).

Using the numerical solutions for Eqs.~\eqref{eq:eq2}, the Eliashberg equation can be viewed as a matrix equation $|\Phi \rangle= \hat K(T,\mu) |{\Phi}\rangle$ (after imposing a large enough cutoff in frequency) and the largest eigenvalue of the kernel $\hat K$ can be computed. {The Eliashberg equation has a nontrivial solution  when the largest eigenvalue reaches 1, indicating the onset of pairing,} and we can map out the boundary of the superconducting region in the $T$-$\mu$ phase diagram. The numerical results for $N=4M$, $m_0=2\omega_0$ and $N=M$, $m_0=2\omega_0$ are shown in Fig.~\ref{fig:fig1} (a) and (b), respectively. The thermal critical point may lie inside or outside the superconducting region depending on the ratio of $M/N$, as discussed above.
\begin{figure}[!htp]
	\centering
	\includegraphics[width=\columnwidth]{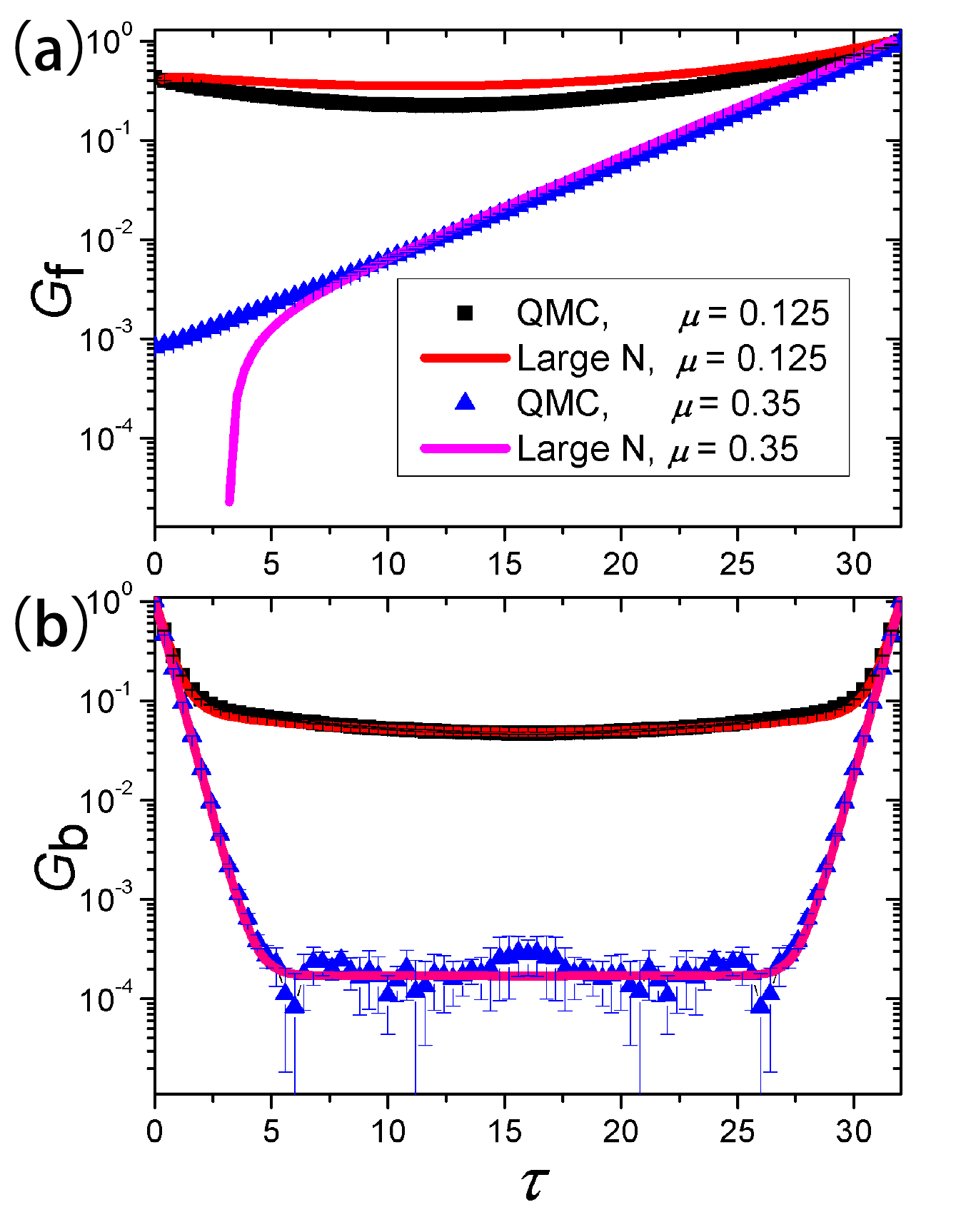}
	\caption{The QMC fermonic Green's functions (a) and the bosonic Green's functions (b) with different $\mu$. $M=4$, $N=16$, $\beta=32$, $\omega_0=1$, $m_0=2$, plot with semi-log axes . For convenience both the $G_f$ and $G_b$ have been normalized to 1 at $\tau=\beta$. The system become gapped with the increase of the chemical potential. The sharp downturn of the large-$N$ result in panel (a) is an artifact of keeping finite frequency points.}
	\label{fig:fig3}
\end{figure}
\begin{figure*}[!htp]
	\includegraphics[width=\textwidth]{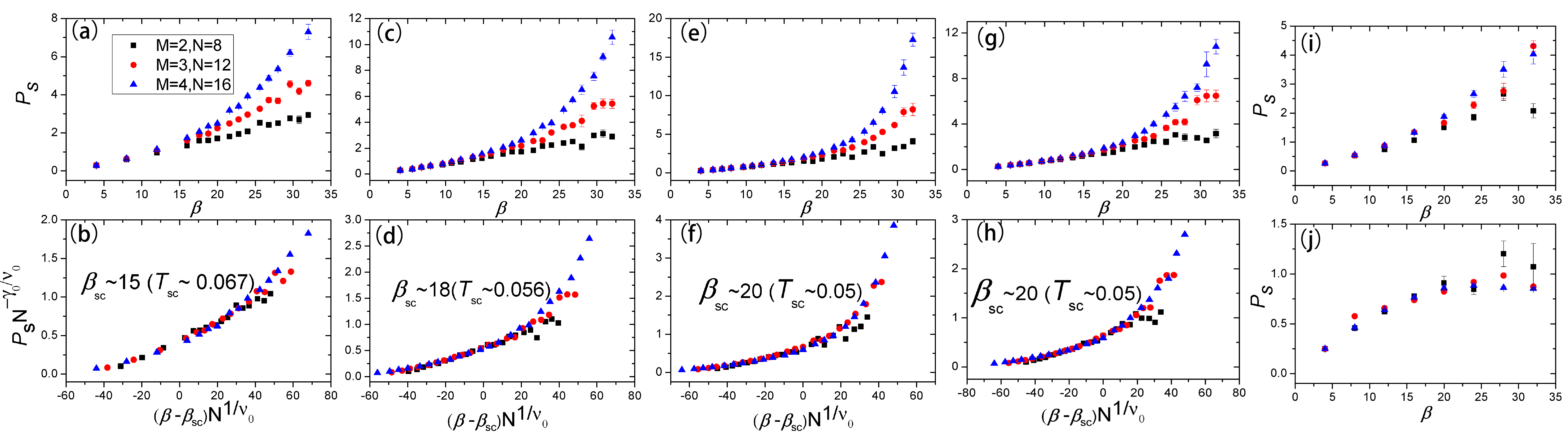}
	\caption{Pair susceptibility $P_s$ measured at different chemical potential $\mu$. The obtained $T_c$ ($\beta_c$) are denoted by the blue triangles in Fig.~\ref{fig:fig1} (a). From the temperature dependence of the $P_s$ with different system sizes $(M,N)$ we perform the data collapse using mean-field exponents $\gamma_0 = 1$, $\nu_0 = 2$ and the transition temperatures $T_c$ ($\beta_c$) are obtained accordingly. The parameters are: $N=4M$, $\omega_0=1$, $m_0=2$. $\mu=0$ in (a) and (b); $\mu=0.05$ in (c) and (d); $\mu=0.125$ in (e) and (f); $\mu=0.2$ in (g) and (h); $\mu=0.25$ in (i); $\mu=0.35$ in (j). The superconducting transition temperature reduces as $\mu$ increases. For $\mu=0.25$ (i) and $\mu=0.35$ (j) the pairing susceptibilities are not divergent at larger $N$, and the system enter a gapped insulator phase.} 
	\label{fig:fig4}
\end{figure*}
\section{ Results from QMC}
To analyze the phase diagram in Fig.~\ref{fig:fig1} with QMC, we focus on the Green's functions and pairing susceptibility obtained in simulations at finite $M,N$. We first perform the QMC simulations at the parameters of $N=4M, \omega_0=1, m_0=2$ with different $\beta\equiv 1/T$ and $\mu$.

We show in Fig.~\ref{fig:fig3} the QMC Green's functions for large $\beta$ with different $\mu$, with $G_{f}(\tau,0)=\frac{1}{(MN)^2}\sum_{i,j=1}^M\sum_{\alpha,\beta=1}^N\langle c_{i,\alpha,\sigma}(\tau)c^\dagger_{j,\beta,\sigma}(0)\rangle$ and $G_{b}(\tau,0)=\frac{1}{N(N-1)}\sum_{\alpha,\beta=1,\alpha\neq\beta}^{N}\langle \phi_{\alpha\beta}(\tau)\phi_{\alpha\beta}(0) \rangle$. One can clearly see that they exhibit distinct behaviors for small and large $\mu$, consistent with the phase diagram in Fig.~\ref{fig:fig1}(a). At $\mu=0.125$, both $G_f$ and $G_b$ decays slowly in imaginary time, similar to the results in Ref.~\cite{2020Yukawa} exhibiting power-law scaling. Note that, for $\mu\neq 0$, since the system is no longer particle-hole symmetric,  $G_f$ is not symmetric with respect to $\tau=\beta/2$, and we normalize the data with respect to $G_f(\tau=\beta)$ and $G_b(\tau=\beta)$. At larger doping, with $\mu=0.35$, both $G_f$ and $G_b$  decay exponentially, consistent with insulating behavior. Since in QMC simulation with finite $N,M$, the system does not develop superconductivity, and it is sensible to compare with Green's functions at large-$N$ for the normal state. As can be seen in Fig.~\ref{fig:fig3}, the agreement is excellent.


For $M=N$ case in Fig.~\ref{fig:fig1} (b), the thermal critical point is located around $(\mu_c=0.3825,T_c=0.07)$, which is within reach of our QMC simulations. We compute the $n (\mu)$ curves (whose derivative is the charge compressibility) near and far away from $T_c$.  We can see in Fig.~\ref{fig:fig2}  that, in excellent agreement with the large-$N$ solution, the compressibility is constant when the temperature is much higher than the critical point ($M=N=8,T=0.125$), while there is a jump in $n (\mu)$ when the temperature is close to the critical point ($M=N=8,10,12,T=0.083$), consistent with the phase diagram in Fig.~\ref{fig:fig1} (b) from large-$N$.

For $N=4M$ our QMC results further reveal that the nFL develop a superconductivity at low temperature in a wide range of chemical potential, reaching beyond the would-be first order phase transition.  To extract the superconducting transition temperature, we measure the pairing correlation in our QMC simulation, and analyze its scaling behavior as system size. The pair susceptibility is expressed as $P_s=\int_0^\beta \text{d}\tau\langle\Delta(\tau)\Delta^\dagger(0)\rangle$, where $\Delta$ is the pairing field defined as $\Delta^\dagger=\frac{1}{\sqrt{MN}}\sum_{i=1}^M\sum_{\alpha=1}^N c_{i,\alpha,\uparrow}^\dagger c_{i,\alpha,\downarrow}^\dagger$. At finite $N$, the pairing susceptibility $P_s$ does not diverge, and can be written as $P_s^{(N)}\sim N^af\left[N^{1/\nu}(T-T_{sc})\right]$, in which $N$ (and $M$ for a fixed ratio) plays the role of the system size ~\cite{isakov2003interplay,Paiva2004,costa2018phonon,ChuangChen2020}.
For our large-$N$ system without the notion of space, the role of correlation length is replaced with a correlation ``cluster size" $n\sim (T-T_{sc})^{-\nu}$, and hence the functional dependence of $f(x)$.  In the large-$N$ limit, all fluctuation effects are suppressed by $1/N$, and such a phase transition is mean-field like~\cite{PhysRevLett.124.017002}. This means that for a fixed $T-T_c$ the exponent $\nu=2$ following the analog of the Josephson’s identity~\footnote{For a mean field theory in finite spatial dimensions, Josephson's identity states that $\nu d =2$, and in our model the dimensionality $d$ does not enter the theory. We have instead $\nu=2$. We thank Ilya Esterlis and Joerg Schmalian for sharing their unpublished results with us on this.}, and that $f(x)\sim 1/x$. Further requiring that in the large-$N$ (thermodynamic) limit the susceptibility diverges independent of $N$,  we obtain that $a=1/2$. Using these exponents, we indeed obtain decent finite size scaling with a $\beta_{sc}$ by data collapse, see Fig.~\ref{fig:fig4} (a)$\sim$(h) with different fermion densities (different $\mu$). We can see that when the $\mu$ increase the superconducting transition temperature is moderately reduced until a sudden drop at larger $\mu$. For $\mu> 0.25$ the pairing susceptibility no longer diverges with large $N$, and the system does not form a pairing state (see Fig.~\ref{fig:fig4}(i) and (j)). The corresponding QMC $T_{sc}$ points are shown in the Fig.~\ref{fig:fig1} (a). The values of $T_{sc}(\mu)$ from QMC are larger than their large-$N$ counterparts, but they are close. In particular, the values of $\mu_{sc}$ from QMC and large-$N$ are in good agreement, consistent with analytical result Eq.~\eqref{eq:musc}.

\section{Discussion}
With combined analytical and numerical efforts, we reveal the $T-\mu$ phase diagram of the spin-1/2 Yukawa-SYK model. We identified that an underlying  first-order quantum phase transition between a non-Fermi liquid and an insulator leads to a dome-like structure of the pairing phase, and depending on the parameter $N/M$, survives at finite-$T$ until a second-order thermal tri-critical point between non-Fermi liquid and an insulator.  The first-order quantum phase transition and the associated thermal critical
point is shared by the original complex SYK model at finite density~\cite{PhysRevLett.120.061602,smit2020quantum}.  In addition,
the superconducting dome in the vicinity of a nFL phase we observed for the spin-1/2 Yukawa-SYK model analytically and numerically in this work is reminiscent of the phase diagrams of many unconventional
superconductors. Our results provide the model realization of the SYK-type nFL and its transition towards superconductivity and insulating states, therefore offer a controlled platform for future investigations of the generic phase diagram that hosts nFL, insulator and superconductor phases and their transitions at generic fermion densities.

It will be interesting to further investigate the scaling behavior of the thermal tri-critical point and determine its universality
class, which we leave this to future work.

\acknowledgments
We thank Andrey Chubukov, Ilya Esterlis, Yingfei Gu, Grigory Tarnopolsky, Joerg Schmalian, Subir Sachdev, Steven Kivelson for insightful discussions. AD and YW are supported by startup funds at the University of Florida. WW, GPP and ZYM acknowledge support from the RGC of Hong Kong SAR of China (Grant Nos. 17303019 and 17301420), MOST through the National Key Research and Development Program (Grant No. 2016YFA0300502) and the Strategic Priority Research Program of the Chinese Academy of Sciences (Grant No. XDB33000000). We thank the Computational Initiative at the Faculty of Science and the Information Technology Services at the University of Hong Kong and Tianhe-2 platform at the National Supercomputer Centers in Guangzhou for their technical support and generous allocation of CPU time.

\begin{appendix}

\section{Model and Quantum Monte Carlo Simulation}
\begin{figure}[!htp]
	\centering
	\includegraphics[width=8cm]{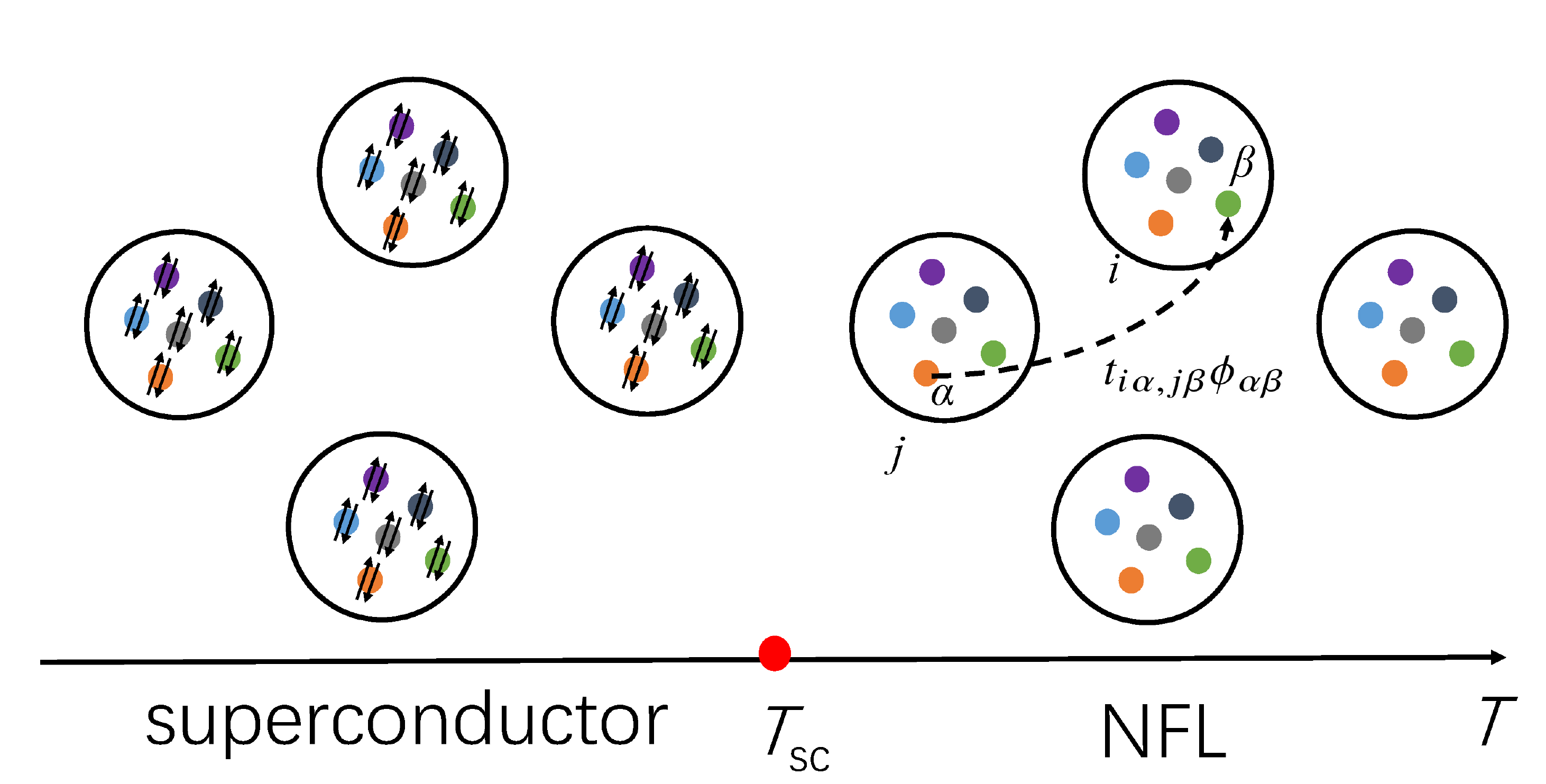}
	\caption{spin-1/2 Yukawa-SYK model. There are $i,j=1,\cdots,M$ quantum dots, each dot acquires $\alpha,\beta=1,\cdots,N$ flavors. Fermions are Yukawa coupled via the random hopping $t_{ij}$ and anti-symmetric bosonic field $\phi_{\alpha \beta}$. The system goes through a phase transition from non-Fermi liquid to a pairing state when the temperature is below the superconducting critical temperature ($T_{sc}$).}
	\label{fig:figs1}
\end{figure}

The Hamiltonian of Eq.~\eqref{eq:eq1} in the main text is illustrated by Fig.~\ref{fig:figs1}. There are $i,j=1,\cdots,M$ quantum dots, each dot acquires $\alpha,\beta=1,\cdots,N$ flavors of fermions. Fermions are Yukawa coupled via the random hopping $t_{ij}$ and anti-symmetric bosonic field $\phi_{\alpha \beta}$. The system go through a phase transition from non-Fermi liquid to a pairing state when the temperature is below the superconducting critical temperature ($T_{sc}$).

We use the DQMC method to simulate this Hamiltonian, and the starting point is the partition function of the system
	 \begin{eqnarray}
\begin{split}
	 	Z=&\operatorname{Tr}\left\{e^{-\beta \hat{H}}\right\} \\
=&\operatorname{Tr}\left\{\left( e^{-\Delta \tau \hat{H}} \right)^{L_{\tau}}\right\} \\
	 	=& \sum_{\{\Phi\}} \operatorname{Tr}_{\mathbf{F}} \langle\Phi^1|e^{-\Delta\tau H}|\Phi^{L_{\tau}}\rangle\langle\Phi^{L_{\tau}}|e^{-\Delta\tau H}|\Phi^{L_{\tau}-1}\rangle\\
&\cdots\langle\Phi^2|e^{-\Delta\tau H}|\Phi^{1}\rangle
	 	\label{eq:S1}
\end{split}
	 \end{eqnarray}
	 where we divide the imaginary time axis $\beta$ into $L_{\tau}$ slices, then we have $\beta = L_{\tau} \times \Delta \tau$. Here
$\Phi_l = \left(\phi_{11,l},\phi_{12,l},\cdots,\phi_{N(N-1),l},\phi_{NN,l}\right)$ is the complete basis of imaginary time propagation in the path integral.
Using Trotter-Suzuki decomposition to the Hamiltonian in Eq.~\eqref{eq:S1},
	 \begin{equation}\label{eq:eq5}
	 	e^{-\Delta \tau \hat{H} }\approx e^{-\Delta \tau \hat{H}_{fb} } e^{-\Delta \tau \hat{H}_{b} }
	 \end{equation}
where $H_{fb}$ is the boson-fermion term, $H_{b}$ is the boson term in the Hamiltonian.
		
Then the partition function can be written as
\begin{align}\label{eq:eq8}
 Z=&\sum_{\{\Phi\}}\omega_B[\phi]\omega_F[\phi].
\end{align}
As for the bosonic part of the partition function,
\small
\begin{equation}\label{eq:eq11}
\begin{aligned}
  \omega_b[\phi]=&C^{L_{\tau}}\left( \prod_{l=1}^{L_{\tau}} \prod_{\alpha , \beta=1}^{N} e^{-\Delta \tau \frac{m_0^2}{2}\phi_{\alpha \beta,l}^2}\right)
  \cdot\left(\prod_{\left<l,l'\right>} \prod_{\alpha , \beta=1}^{N} e^{-\frac{\left( \phi_{\alpha \beta,l}-\phi_{\alpha \beta,l'}\right)^2}{2 \Delta \tau}}\right)
\end{aligned}
\end{equation}
\normalsize
with $\left<l,l'\right> $ stands for the nearest-neighbor interaction in imaginary time direction, and $C$ is a constant. As for the fermionic part of the partition function
\begin{align}\label{eq:eq13}
 \omega_F[\phi]=\det[I+\mathbf{B}^{L_{\tau}}\mathbf{B}^{L_{\tau}-1}\cdots \mathbf{B}^l\cdots \mathbf{B}^2\mathbf{B}^1],
\end{align}
where
	 \begin{equation}\label{eq:eq14}
	 B^l= e^{-\Delta \tau V(\Phi_l) }
	 \end{equation}
and

	\begin{equation}\label{eq:eq15}
	\begin{aligned}
	V(\Phi_l)=&\frac{i}{\sqrt{MN}}  \sigma^z_{2\times 2}\otimes \left( t_{i j}\right)_{M \times M} \otimes \left( \phi_{\alpha \beta,l} \right)_{N \times N}-\mu\otimes \mathbb{I}_{2MN\times 2MN}
	\end{aligned}
	\end{equation}
Here $\mathbb{I}$ is  identity matrix.

With these notations prepared, finally the partition function in Eq.~\eqref{eq:eq8} can be written as

	\begin{equation} \begin{aligned}\label{eq:eq16}
	Z=&\sum_{\{\Phi\}} \prod_{l=1}^{L_{\tau}}  C^{L_{\tau}}\left( \prod_{l=1}^{L_{\tau}} \prod_{\alpha , \beta=1}^{N} e^{-\Delta \tau \frac{M}{N}\frac{m_0^2}{2}\phi_{\alpha \beta,l}^2}  \right)\cdot\\
 &\left( \prod_{\left<l,l'\right>} \prod_{\alpha , \beta=1}^{N} e^{-\frac{\left( \phi_{\alpha \beta,l}-\phi_{\alpha \beta,l'}\right)^2}{2 \Delta \tau}}  \right)\cdot\\
	&\operatorname{Det}\left[\mathbf{I}+\mathbf{B}^{L_{\tau}}\mathbf{B}^{L_{\tau}-1}\cdots \mathbf{B}^l\cdots \mathbf{B}^2\mathbf{B}^1\right]
	\end{aligned}
	\end{equation}

This partition function
is free from the minus-sign problem with any $\mu$. For the part of boson-fermion term of the Hamiltonian, it is
invariant under a time-reversal  symmetry operation $\mathcal{T}=i\sigma_y \mathcal{K}$. Here the $\mathcal{K}$ is the complex conjugate operator.
The boson-fermion term of the Hamiltonian can be written as
\begin{equation}
\begin{aligned}
\hat H_{fb}=&\sum_{ij=1}^{N}\sum_{\alpha,\beta=1}^{N}\frac{i}{\sqrt{MN}}t_{\alpha\beta}\phi_{ij}c_{i\alpha\uparrow}^\dagger c_{j\alpha\uparrow}-\mu c_{i\alpha\uparrow}^\dagger c_{i\alpha\uparrow}\\ &-\frac{i}{\sqrt{MN}}t_{\alpha\beta}\phi_{ij}c_{i\alpha\downarrow}^\dagger c_{j\alpha\downarrow}-\mu c_{i\alpha\downarrow}^\dagger c_{i\alpha\downarrow}
\end{aligned}
\end{equation}
\begin{figure}[htp!]
	\centering
	\includegraphics[width=8cm]{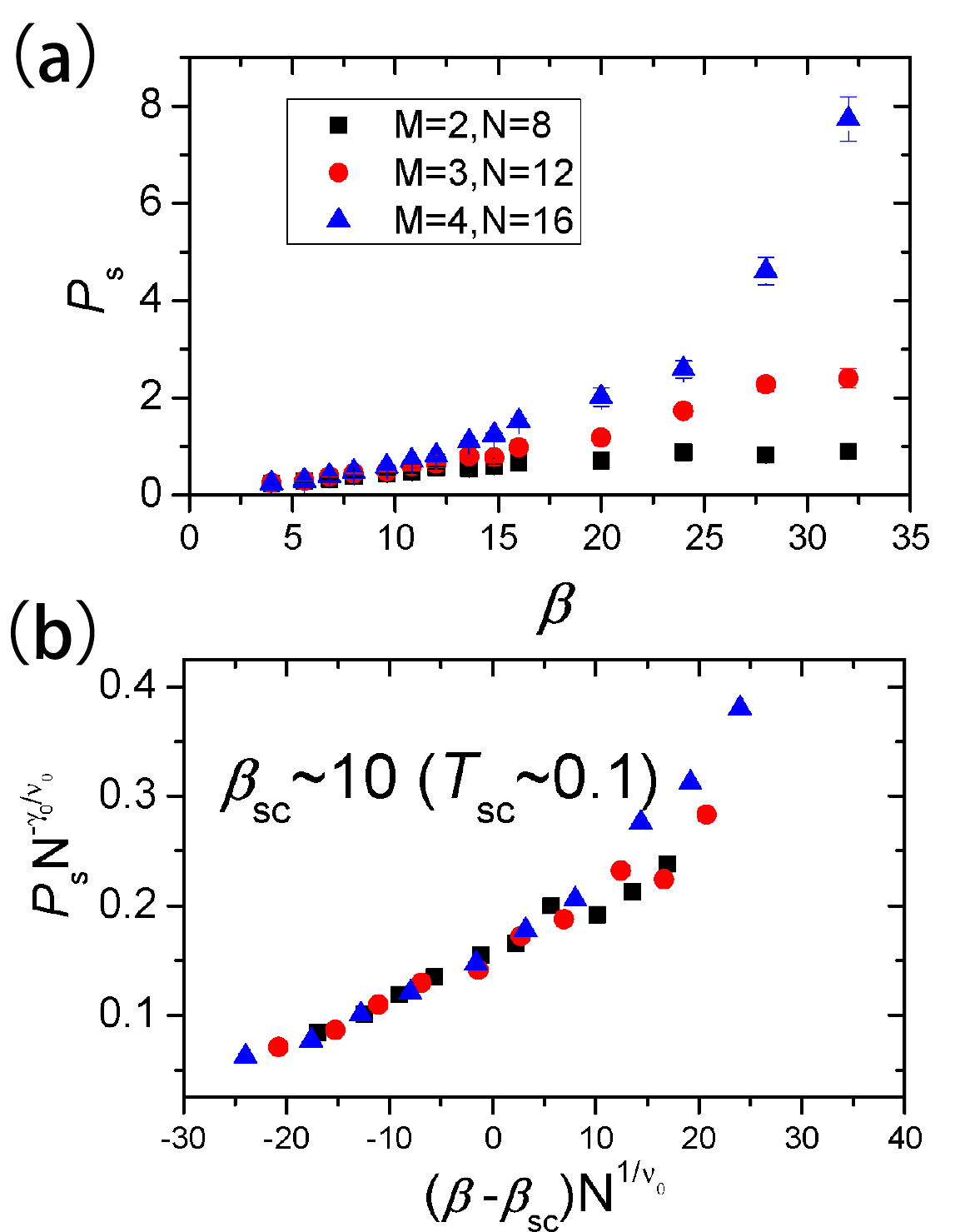}
	\caption{Pair susceptibility for different $\beta$ and data collapse, $\omega_0=1$, $m_0=1$, $\mu=0$. The superconducting transition temperature is around $T_{sc}\sim0.1$ $(\beta_c\sim10)$, which is higher than the superconducting transition temperature around $T_{sc}\sim0.067 \ (\beta_{sc}\sim15)$ at $\omega_0=1, m_0=2$ and $\mu=0$.}
	\label{fig:figS2}
\end{figure}
Under the transformation of $\mathcal{T}=i\sigma_y\mathcal{K}$, we have
\begin{equation} \begin{aligned}
\mathcal{T} H_{fb} \mathcal{T}^{-1}=&\sum_{ij=1}^{N}\sum_{\alpha,\beta=1}^{N}-\frac{i}{\sqrt{MN}}t_{\alpha\beta}\phi_{ij}c_{i\alpha\downarrow}^\dagger c_{j\alpha\downarrow}-\mu c_{i\alpha\downarrow}^\dagger c_{i\alpha\downarrow}\\
&+\frac{i}{\sqrt{MN}}t_{\alpha\beta}\phi_{ij}c_{i\alpha\uparrow}^\dagger c_{j\alpha\uparrow}-\mu c_{i\alpha\uparrow}^\dagger c_{i\alpha\uparrow}\\
=&H_{fb}
\end{aligned}
\end{equation}
At the same time for the fermion determinant
\begin{eqnarray}
&&\det\left[1+B(\beta,0)\right] \nonumber\\
&=&\det\left[1+B^\uparrow(\beta,0)\right]\det\left[1+B^\downarrow(\beta,0)\right]\nonumber\\
&=&\det\left[1+B^\uparrow(\beta,0)\right]\det\left[\mathcal{T}\left(1+B^\downarrow(\beta,0)\right)\mathcal{T}^{-1}\right]^*\nonumber\\
&=&\det\left[1+B^\uparrow(\beta,0)\right]\det\left[1+B^\uparrow(\beta,0)\right]^*\\
&=&\left| \det\left[1+B^\uparrow(\beta,0)\right]\right| ^2
\end{eqnarray}
where $B(\beta,0) =\mathbf{B}^{L_{\tau}}\mathbf{B}^{L_{\tau}-1}\cdots \mathbf{B}^l\cdots \mathbf{B}^2\mathbf{B}^1$. The determinant is a positive and real number. Also for the boson
part of the weight $\omega_b[\phi]$ is positive and real. Therefore we have proved that this Hamiltonian is sign problem free.

From a simpler viewpoint, just by looking at the matrix elements of matrices corresponding to different spins: $B^\uparrow(\beta,0)$ and $B^\downarrow(\beta,0)$, we can see this model doesn't have sign problem. Every element of $1+B^\downarrow(\beta,0)$ is individually complex conjugate with the corresponding element of $1+B^\uparrow(\beta,0)$, which means :
\begin{equation}
\det\left[1+B^\uparrow(\beta,0)\right]=\det\left[1+B^\downarrow(\beta,0)\right]^*
\end{equation}

\section{The critical temperature of superconducting with different $m_0$ }	
 The influence of the ratio $\omega_0/m_0$ to the superconducting transition temperature has been studied quantitativly by large-$N$ limit calculation. The inverse transition temperature $\beta_c$ from nFL to superconductivity as a function of the ratio $\omega_0/m_0$ for $N = 4M$
and $N = M$ are discussed in the main text here and in the Ref.~\cite{2020Yukawa}.  By QMC simulation ,we also obtained that when $m_0=1,\ \omega_0=1$ and $\mu=0$, the superconducting transition temperature is around $T_{sc}\sim0.1$ $(\beta_{sc}\sim10)$, which is higher than the superconducting transition temperature around $T_{sc}\sim  0.067 \ (\beta_{sc} \sim 15)$ at $\omega_0=1, m_0=2$ and $\mu=0$. The results are shown in Fig.~\ref{fig:figS2} and consistent with theoretical analysis in Ref.~\cite{2020Yukawa}.

\end{appendix}
\bibliography{qmc-Yukawa-SYK-SC}

\begin{thebibliography}{40}%
\makeatletter
\providecommand \@ifxundefined [1]{%
 \@ifx{#1\undefined}
}%
\providecommand \@ifnum [1]{%
 \ifnum #1\expandafter \@firstoftwo
 \else \expandafter \@secondoftwo
 \fi
}%
\providecommand \@ifx [1]{%
 \ifx #1\expandafter \@firstoftwo
 \else \expandafter \@secondoftwo
 \fi
}%
\providecommand \natexlab [1]{#1}%
\providecommand \enquote  [1]{``#1''}%
\providecommand \bibnamefont  [1]{#1}%
\providecommand \bibfnamefont [1]{#1}%
\providecommand \citenamefont [1]{#1}%
\providecommand \href@noop [0]{\@secondoftwo}%
\providecommand \href [0]{\begingroup \@sanitize@url \@href}%
\providecommand \@href[1]{\@@startlink{#1}\@@href}%
\providecommand \@@href[1]{\endgroup#1\@@endlink}%
\providecommand \@sanitize@url [0]{\catcode `\\12\catcode `\$12\catcode
  `\&12\catcode `\#12\catcode `\^12\catcode `\_12\catcode `\%12\relax}%
\providecommand \@@startlink[1]{}%
\providecommand \@@endlink[0]{}%
\providecommand \url  [0]{\begingroup\@sanitize@url \@url }%
\providecommand \@url [1]{\endgroup\@href {#1}{\urlprefix }}%
\providecommand \urlprefix  [0]{URL }%
\providecommand \Eprint [0]{\href }%
\providecommand \doibase [0]{http://dx.doi.org/}%
\providecommand \selectlanguage [0]{\@gobble}%
\providecommand \bibinfo  [0]{\@secondoftwo}%
\providecommand \bibfield  [0]{\@secondoftwo}%
\providecommand \translation [1]{[#1]}%
\providecommand \BibitemOpen [0]{}%
\providecommand \bibitemStop [0]{}%
\providecommand \bibitemNoStop [0]{.\EOS\space}%
\providecommand \EOS [0]{\spacefactor3000\relax}%
\providecommand \BibitemShut  [1]{\csname bibitem#1\endcsname}%
\let\auto@bib@innerbib\@empty
\bibitem [{\citenamefont {Wang}\ and\ \citenamefont
  {Chubukov}(2020)}]{PhysRevResearch.2.033084}%
  \BibitemOpen
  \bibfield  {author} {\bibinfo {author} {\bibfnamefont {Yuxuan}\ \bibnamefont
  {Wang}}\ and\ \bibinfo {author} {\bibfnamefont {Andrey~V.}\ \bibnamefont
  {Chubukov}},\ }\bibfield  {title} {\enquote {\bibinfo {title} {Quantum phase
  transition in the {Yukawa-SYK} model},}\ }\href {\doibase
  10.1103/PhysRevResearch.2.033084} {\bibfield  {journal} {\bibinfo  {journal}
  {Phys. Rev. Research}\ }\textbf {\bibinfo {volume} {2}},\ \bibinfo {pages}
  {033084} (\bibinfo {year} {2020})}\BibitemShut {NoStop}%
\bibitem [{\citenamefont {Keimer}\ \emph {et~al.}(2015)\citenamefont {Keimer},
  \citenamefont {Kivelson}, \citenamefont {Norman}, \citenamefont {Uchida},\
  and\ \citenamefont {Zaanen}}]{Keimer2015}%
  \BibitemOpen
  \bibfield  {author} {\bibinfo {author} {\bibfnamefont {B.}~\bibnamefont
  {Keimer}}, \bibinfo {author} {\bibfnamefont {S.~A.}\ \bibnamefont
  {Kivelson}}, \bibinfo {author} {\bibfnamefont {M.~R.}\ \bibnamefont
  {Norman}}, \bibinfo {author} {\bibfnamefont {S.}~\bibnamefont {Uchida}}, \
  and\ \bibinfo {author} {\bibfnamefont {J.}~\bibnamefont {Zaanen}},\
  }\bibfield  {title} {\enquote {\bibinfo {title} {From quantum matter to
  high-temperature superconductivity in copper oxides},}\ }\href {\doibase
  10.1038/nature14165} {\bibfield  {journal} {\bibinfo  {journal} {Nature}\
  }\textbf {\bibinfo {volume} {518}},\ \bibinfo {pages} {179--186} (\bibinfo
  {year} {2015})}\BibitemShut {NoStop}%
\bibitem [{\citenamefont {Liu}\ \emph {et~al.}(2016)\citenamefont {Liu},
  \citenamefont {Gu}, \citenamefont {Zhang}, \citenamefont {Gong},
  \citenamefont {Zhang}, \citenamefont {Xie}, \citenamefont {Lu}, \citenamefont
  {Ma}, \citenamefont {Zhang}, \citenamefont {Zhang}, \citenamefont {Zhu},
  \citenamefont {Ren}, \citenamefont {Shan}, \citenamefont {Qiu}, \citenamefont
  {Dai}, \citenamefont {Yang}, \citenamefont {Luo},\ and\ \citenamefont
  {Li}}]{ZhaoyuLiu2016}%
  \BibitemOpen
  \bibfield  {author} {\bibinfo {author} {\bibfnamefont {Zhaoyu}\ \bibnamefont
  {Liu}}, \bibinfo {author} {\bibfnamefont {Yanhong}\ \bibnamefont {Gu}},
  \bibinfo {author} {\bibfnamefont {Wei}\ \bibnamefont {Zhang}}, \bibinfo
  {author} {\bibfnamefont {Dongliang}\ \bibnamefont {Gong}}, \bibinfo {author}
  {\bibfnamefont {Wenliang}\ \bibnamefont {Zhang}}, \bibinfo {author}
  {\bibfnamefont {Tao}\ \bibnamefont {Xie}}, \bibinfo {author} {\bibfnamefont
  {Xingye}\ \bibnamefont {Lu}}, \bibinfo {author} {\bibfnamefont {Xiaoyan}\
  \bibnamefont {Ma}}, \bibinfo {author} {\bibfnamefont {Xiaotian}\ \bibnamefont
  {Zhang}}, \bibinfo {author} {\bibfnamefont {Rui}\ \bibnamefont {Zhang}},
  \bibinfo {author} {\bibfnamefont {Jun}\ \bibnamefont {Zhu}}, \bibinfo
  {author} {\bibfnamefont {Cong}\ \bibnamefont {Ren}}, \bibinfo {author}
  {\bibfnamefont {Lei}\ \bibnamefont {Shan}}, \bibinfo {author} {\bibfnamefont
  {Xianggang}\ \bibnamefont {Qiu}}, \bibinfo {author} {\bibfnamefont
  {Pengcheng}\ \bibnamefont {Dai}}, \bibinfo {author} {\bibfnamefont {Yi-feng}\
  \bibnamefont {Yang}}, \bibinfo {author} {\bibfnamefont {Huiqian}\
  \bibnamefont {Luo}}, \ and\ \bibinfo {author} {\bibfnamefont {Shiliang}\
  \bibnamefont {Li}},\ }\bibfield  {title} {\enquote {\bibinfo {title}
  {{Nematic Quantum Critical Fluctuations in
  ${\mathrm{BaFe}}_{2\ensuremath{-}x}{\mathrm{Ni}}_{x}{\mathrm{As}}_{2}$}},}\
  }\href {\doibase 10.1103/PhysRevLett.117.157002} {\bibfield  {journal}
  {\bibinfo  {journal} {Phys. Rev. Lett.}\ }\textbf {\bibinfo {volume} {117}},\
  \bibinfo {pages} {157002} (\bibinfo {year} {2016})}\BibitemShut {NoStop}%
\bibitem [{\citenamefont {Gu}\ \emph {et~al.}(2017{\natexlab{a}})\citenamefont
  {Gu}, \citenamefont {Liu}, \citenamefont {Xie}, \citenamefont {Zhang},
  \citenamefont {Gong}, \citenamefont {Hu}, \citenamefont {Ma}, \citenamefont
  {Li}, \citenamefont {Zhao}, \citenamefont {Lin}, \citenamefont {Xu},
  \citenamefont {Tan}, \citenamefont {Chen}, \citenamefont {Meng},
  \citenamefont {Yang}, \citenamefont {Luo},\ and\ \citenamefont
  {Li}}]{YanhongGu2017}%
  \BibitemOpen
  \bibfield  {author} {\bibinfo {author} {\bibfnamefont {Yanhong}\ \bibnamefont
  {Gu}}, \bibinfo {author} {\bibfnamefont {Zhaoyu}\ \bibnamefont {Liu}},
  \bibinfo {author} {\bibfnamefont {Tao}\ \bibnamefont {Xie}}, \bibinfo
  {author} {\bibfnamefont {Wenliang}\ \bibnamefont {Zhang}}, \bibinfo {author}
  {\bibfnamefont {Dongliang}\ \bibnamefont {Gong}}, \bibinfo {author}
  {\bibfnamefont {Ding}\ \bibnamefont {Hu}}, \bibinfo {author} {\bibfnamefont
  {Xiaoyan}\ \bibnamefont {Ma}}, \bibinfo {author} {\bibfnamefont {Chunhong}\
  \bibnamefont {Li}}, \bibinfo {author} {\bibfnamefont {Lingxiao}\ \bibnamefont
  {Zhao}}, \bibinfo {author} {\bibfnamefont {Lifang}\ \bibnamefont {Lin}},
  \bibinfo {author} {\bibfnamefont {Zhuang}\ \bibnamefont {Xu}}, \bibinfo
  {author} {\bibfnamefont {Guotai}\ \bibnamefont {Tan}}, \bibinfo {author}
  {\bibfnamefont {Genfu}\ \bibnamefont {Chen}}, \bibinfo {author}
  {\bibfnamefont {Zi~Yang}\ \bibnamefont {Meng}}, \bibinfo {author}
  {\bibfnamefont {Yi-feng}\ \bibnamefont {Yang}}, \bibinfo {author}
  {\bibfnamefont {Huiqian}\ \bibnamefont {Luo}}, \ and\ \bibinfo {author}
  {\bibfnamefont {Shiliang}\ \bibnamefont {Li}},\ }\bibfield  {title} {\enquote
  {\bibinfo {title} {Unified phase diagram for iron-based superconductors},}\
  }\href {\doibase 10.1103/PhysRevLett.119.157001} {\bibfield  {journal}
  {\bibinfo  {journal} {Phys. Rev. Lett.}\ }\textbf {\bibinfo {volume} {119}},\
  \bibinfo {pages} {157001} (\bibinfo {year} {2017}{\natexlab{a}})}\BibitemShut
  {NoStop}%
\bibitem [{\citenamefont {Custers}\ \emph {et~al.}(2003)\citenamefont
  {Custers}, \citenamefont {Gegenwart}, \citenamefont {Wilhelm}, \citenamefont
  {Neumaier}, \citenamefont {Tokiwa}, \citenamefont {Trovarelli}, \citenamefont
  {Geibel}, \citenamefont {Steglich}, \citenamefont {Pépin},\ and\
  \citenamefont {Coleman}}]{Custers2003}%
  \BibitemOpen
  \bibfield  {author} {\bibinfo {author} {\bibfnamefont {J.}~\bibnamefont
  {Custers}}, \bibinfo {author} {\bibfnamefont {P.}~\bibnamefont {Gegenwart}},
  \bibinfo {author} {\bibfnamefont {H.}~\bibnamefont {Wilhelm}}, \bibinfo
  {author} {\bibfnamefont {K.}~\bibnamefont {Neumaier}}, \bibinfo {author}
  {\bibfnamefont {Y.}~\bibnamefont {Tokiwa}}, \bibinfo {author} {\bibfnamefont
  {O.}~\bibnamefont {Trovarelli}}, \bibinfo {author} {\bibfnamefont
  {C.}~\bibnamefont {Geibel}}, \bibinfo {author} {\bibfnamefont
  {F.}~\bibnamefont {Steglich}}, \bibinfo {author} {\bibfnamefont
  {C.}~\bibnamefont {Pépin}}, \ and\ \bibinfo {author} {\bibfnamefont
  {P.}~\bibnamefont {Coleman}},\ }\bibfield  {title} {\enquote {\bibinfo
  {title} {The break-up of heavy electrons at a quantum critical point},}\
  }\href {\doibase 10.1038/nature01774} {\bibfield  {journal} {\bibinfo
  {journal} {Nature}\ }\textbf {\bibinfo {volume} {424}},\ \bibinfo {pages}
  {524--527} (\bibinfo {year} {2003})}\BibitemShut {NoStop}%
\bibitem [{\citenamefont {Shen}\ \emph
  {et~al.}(2020{\natexlab{a}})\citenamefont {Shen}, \citenamefont {Zhang},
  \citenamefont {Komijani}, \citenamefont {Nicklas}, \citenamefont {Borth},
  \citenamefont {Wang}, \citenamefont {Chen}, \citenamefont {Nie},
  \citenamefont {Li}, \citenamefont {Lu}, \citenamefont {Lee}, \citenamefont
  {Smidman}, \citenamefont {Steglich}, \citenamefont {Coleman},\ and\
  \citenamefont {Yuan}}]{BinShen2019}%
  \BibitemOpen
  \bibfield  {author} {\bibinfo {author} {\bibfnamefont {Bin}\ \bibnamefont
  {Shen}}, \bibinfo {author} {\bibfnamefont {Yongjun}\ \bibnamefont {Zhang}},
  \bibinfo {author} {\bibfnamefont {Yashar}\ \bibnamefont {Komijani}}, \bibinfo
  {author} {\bibfnamefont {Michael}\ \bibnamefont {Nicklas}}, \bibinfo {author}
  {\bibfnamefont {Robert}\ \bibnamefont {Borth}}, \bibinfo {author}
  {\bibfnamefont {An}~\bibnamefont {Wang}}, \bibinfo {author} {\bibfnamefont
  {Ye}~\bibnamefont {Chen}}, \bibinfo {author} {\bibfnamefont {Zhiyong}\
  \bibnamefont {Nie}}, \bibinfo {author} {\bibfnamefont {Rui}\ \bibnamefont
  {Li}}, \bibinfo {author} {\bibfnamefont {Xin}\ \bibnamefont {Lu}}, \bibinfo
  {author} {\bibfnamefont {Hanoh}\ \bibnamefont {Lee}}, \bibinfo {author}
  {\bibfnamefont {Michael}\ \bibnamefont {Smidman}}, \bibinfo {author}
  {\bibfnamefont {Frank}\ \bibnamefont {Steglich}}, \bibinfo {author}
  {\bibfnamefont {Piers}\ \bibnamefont {Coleman}}, \ and\ \bibinfo {author}
  {\bibfnamefont {Huiqiu}\ \bibnamefont {Yuan}},\ }\bibfield  {title} {\enquote
  {\bibinfo {title} {Strange-metal behaviour in a pure ferromagnetic kondo
  lattice},}\ }\href {\doibase 10.1038/s41586-020-2052-z} {\bibfield  {journal}
  {\bibinfo  {journal} {Nature}\ }\textbf {\bibinfo {volume} {579}},\ \bibinfo
  {pages} {51 -- 55} (\bibinfo {year} {2020}{\natexlab{a}})}\BibitemShut
  {NoStop}%
\bibitem [{\citenamefont {Cao}\ \emph {et~al.}(2020)\citenamefont {Cao},
  \citenamefont {Chowdhury}, \citenamefont {Rodan-Legrain}, \citenamefont
  {Rubies-Bigorda}, \citenamefont {Watanabe}, \citenamefont {Taniguchi},
  \citenamefont {Senthil},\ and\ \citenamefont {Jarillo-Herrero}}]{YCao2019}%
  \BibitemOpen
  \bibfield  {author} {\bibinfo {author} {\bibfnamefont {Yuan}\ \bibnamefont
  {Cao}}, \bibinfo {author} {\bibfnamefont {Debanjan}\ \bibnamefont
  {Chowdhury}}, \bibinfo {author} {\bibfnamefont {Daniel}\ \bibnamefont
  {Rodan-Legrain}}, \bibinfo {author} {\bibfnamefont {Oriol}\ \bibnamefont
  {Rubies-Bigorda}}, \bibinfo {author} {\bibfnamefont {Kenji}\ \bibnamefont
  {Watanabe}}, \bibinfo {author} {\bibfnamefont {Takashi}\ \bibnamefont
  {Taniguchi}}, \bibinfo {author} {\bibfnamefont {T.}~\bibnamefont {Senthil}},
  \ and\ \bibinfo {author} {\bibfnamefont {Pablo}\ \bibnamefont
  {Jarillo-Herrero}},\ }\bibfield  {title} {\enquote {\bibinfo {title} {Strange
  metal in magic-angle graphene with near planckian dissipation},}\ }\href
  {\doibase 10.1103/PhysRevLett.124.076801} {\bibfield  {journal} {\bibinfo
  {journal} {Phys. Rev. Lett.}\ }\textbf {\bibinfo {volume} {124}},\ \bibinfo
  {pages} {076801} (\bibinfo {year} {2020})}\BibitemShut {NoStop}%
\bibitem [{\citenamefont {Shen}\ \emph
  {et~al.}(2020{\natexlab{b}})\citenamefont {Shen}, \citenamefont {Chu},
  \citenamefont {Wu}, \citenamefont {Li}, \citenamefont {Wang}, \citenamefont
  {Zhao}, \citenamefont {Tang}, \citenamefont {Liu}, \citenamefont {Tian},
  \citenamefont {Watanabe}, \citenamefont {Taniguchi}, \citenamefont {Yang},
  \citenamefont {Meng}, \citenamefont {Shi}, \citenamefont {Yazyev},\ and\
  \citenamefont {Zhang}}]{shen2019observation}%
  \BibitemOpen
  \bibfield  {author} {\bibinfo {author} {\bibfnamefont {Cheng}\ \bibnamefont
  {Shen}}, \bibinfo {author} {\bibfnamefont {Yanbang}\ \bibnamefont {Chu}},
  \bibinfo {author} {\bibfnamefont {QuanSheng}\ \bibnamefont {Wu}}, \bibinfo
  {author} {\bibfnamefont {Na}~\bibnamefont {Li}}, \bibinfo {author}
  {\bibfnamefont {Shuopei}\ \bibnamefont {Wang}}, \bibinfo {author}
  {\bibfnamefont {Yanchong}\ \bibnamefont {Zhao}}, \bibinfo {author}
  {\bibfnamefont {Jian}\ \bibnamefont {Tang}}, \bibinfo {author} {\bibfnamefont
  {Jieying}\ \bibnamefont {Liu}}, \bibinfo {author} {\bibfnamefont {Jinpeng}\
  \bibnamefont {Tian}}, \bibinfo {author} {\bibfnamefont {Kenji}\ \bibnamefont
  {Watanabe}}, \bibinfo {author} {\bibfnamefont {Takashi}\ \bibnamefont
  {Taniguchi}}, \bibinfo {author} {\bibfnamefont {Rong}\ \bibnamefont {Yang}},
  \bibinfo {author} {\bibfnamefont {Zi~Yang}\ \bibnamefont {Meng}}, \bibinfo
  {author} {\bibfnamefont {Dongxia}\ \bibnamefont {Shi}}, \bibinfo {author}
  {\bibfnamefont {Oleg~V.}\ \bibnamefont {Yazyev}}, \ and\ \bibinfo {author}
  {\bibfnamefont {Guangyu}\ \bibnamefont {Zhang}},\ }\bibfield  {title}
  {\enquote {\bibinfo {title} {Correlated states in twisted double bilayer
  graphene},}\ }\href {\doibase 10.1038/s41567-020-0825-9} {\bibfield
  {journal} {\bibinfo  {journal} {Nature Physics}\ } (\bibinfo {year}
  {2020}{\natexlab{b}}),\ 10.1038/s41567-020-0825-9}\BibitemShut {NoStop}%
\bibitem [{\citenamefont {{Chen}}\ \emph {et~al.}(2020)\citenamefont {{Chen}},
  \citenamefont {{Yuan}}, \citenamefont {{Qi}},\ and\ \citenamefont
  {{Meng}}}]{ChuangChen2020}%
  \BibitemOpen
  \bibfield  {author} {\bibinfo {author} {\bibfnamefont {Chuang}\ \bibnamefont
  {{Chen}}}, \bibinfo {author} {\bibfnamefont {Tian}\ \bibnamefont {{Yuan}}},
  \bibinfo {author} {\bibfnamefont {Yang}\ \bibnamefont {{Qi}}}, \ and\
  \bibinfo {author} {\bibfnamefont {Zi~Yang}\ \bibnamefont {{Meng}}},\
  }\bibfield  {title} {\enquote {\bibinfo {title} {{Doped Orthogonal Metals
  Become Fermi Arcs}},}\ }\href@noop {} {\bibfield  {journal} {\bibinfo
  {journal} {arXiv e-prints}\ ,\ \bibinfo {eid} {arXiv:2007.05543}} (\bibinfo
  {year} {2020})},\ \Eprint {http://arxiv.org/abs/2007.05543} {arXiv:2007.05543
  [cond-mat.str-el]} \BibitemShut {NoStop}%
\bibitem [{\citenamefont {Sachdev}\ and\ \citenamefont {Ye}(2015)}]{SY}%
  \BibitemOpen
  \bibfield  {author} {\bibinfo {author} {\bibfnamefont {Subir}\ \bibnamefont
  {Sachdev}}\ and\ \bibinfo {author} {\bibfnamefont {Jinwu}\ \bibnamefont
  {Ye}},\ }\bibfield  {title} {\enquote {\bibinfo {title} {{Gapless spin-fluid
  ground state in a random quantum Heisenberg magnet}},}\ }\href {\doibase
  10.1103/PhysRevLett.70.3339} {\bibfield  {journal} {\bibinfo  {journal}
  {Phys. Rev. Lett.}\ }\textbf {\bibinfo {volume} {70}},\ \bibinfo {pages}
  {3339--3342} (\bibinfo {year} {2015})}\BibitemShut {NoStop}%
\bibitem [{\citenamefont {Kitaev}()}]{K}%
  \BibitemOpen
  \bibfield  {author} {\bibinfo {author} {\bibfnamefont {A}~\bibnamefont
  {Kitaev}},\ }\bibfield  {title} {\enquote {\bibinfo {title} {Talks at kitp,
  university of california, santa barbara},}\ }\href
  {http://online.kitp.ucsb.edu/online/entangled15/} {\bibinfo  {journal}
  {Entanglement in Strongly-Correlated Quantum Matter}\ }\BibitemShut {NoStop}%
\bibitem [{\citenamefont {Sachdev}(2015)}]{SYK2}%
  \BibitemOpen
\bibfield  {journal} {  }\bibfield  {author} {\bibinfo {author} {\bibfnamefont
  {Subir}\ \bibnamefont {Sachdev}},\ }\bibfield  {title} {\enquote {\bibinfo
  {title} {Bekenstein-hawking entropy and strange metals},}\ }\href {\doibase
  10.1103/PhysRevX.5.041025} {\bibfield  {journal} {\bibinfo  {journal} {Phys.
  Rev. X}\ }\textbf {\bibinfo {volume} {5}},\ \bibinfo {pages} {041025}
  (\bibinfo {year} {2015})}\BibitemShut {NoStop}%
\bibitem [{\citenamefont {Kitaev}\ and\ \citenamefont {Suh}(2018)}]{SYK3}%
  \BibitemOpen
  \bibfield  {author} {\bibinfo {author} {\bibfnamefont {Alexei}\ \bibnamefont
  {Kitaev}}\ and\ \bibinfo {author} {\bibfnamefont {S.~Josephine}\ \bibnamefont
  {Suh}},\ }\bibfield  {title} {\enquote {\bibinfo {title} {The soft mode in
  the sachdev-ye-kitaev model and its gravity dual},}\ }\href {\doibase
  10.1007/JHEP05(2018)183} {\bibfield  {journal} {\bibinfo  {journal} {Journal
  of High Energy Physics}\ }\textbf {\bibinfo {volume} {2018}},\ \bibinfo
  {pages} {183} (\bibinfo {year} {2018})}\BibitemShut {NoStop}%
\bibitem [{\citenamefont {Abanov}\ \emph {et~al.}(2003)\citenamefont {Abanov},
  \citenamefont {Chubukov},\ and\ \citenamefont {Schmalian}}]{abanov2003}%
  \BibitemOpen
  \bibfield  {author} {\bibinfo {author} {\bibfnamefont {A.}~\bibnamefont
  {Abanov}}, \bibinfo {author} {\bibfnamefont {A.}~\bibnamefont {Chubukov}}, \
  and\ \bibinfo {author} {\bibfnamefont {J.}~\bibnamefont {Schmalian}},\
  }\bibfield  {title} {\enquote {\bibinfo {title} {Quantum-critical theory of
  the spin-fermion model and its application to cuprates: normal state
  analysis},}\ }\href@noop {} {\bibfield  {journal} {\bibinfo  {journal}
  {Advances in Physics}\ }\textbf {\bibinfo {volume} {52}},\ \bibinfo {pages}
  {119--218} (\bibinfo {year} {2003})}\BibitemShut {NoStop}%
\bibitem [{\citenamefont {Metlitski}\ and\ \citenamefont
  {Sachdev}(2010{\natexlab{a}})}]{Metlitski2010a}%
  \BibitemOpen
  \bibfield  {author} {\bibinfo {author} {\bibfnamefont {Max~A.}\ \bibnamefont
  {Metlitski}}\ and\ \bibinfo {author} {\bibfnamefont {Subir}\ \bibnamefont
  {Sachdev}},\ }\bibfield  {title} {\enquote {\bibinfo {title} {Quantum phase
  transitions of metals in two spatial dimensions. i. ising-nematic order},}\
  }\href {\doibase 10.1103/PhysRevB.82.075127} {\bibfield  {journal} {\bibinfo
  {journal} {Phys. Rev. B}\ }\textbf {\bibinfo {volume} {82}},\ \bibinfo
  {pages} {075127} (\bibinfo {year} {2010}{\natexlab{a}})}\BibitemShut
  {NoStop}%
\bibitem [{\citenamefont {Metlitski}\ and\ \citenamefont
  {Sachdev}(2010{\natexlab{b}})}]{Metlitski2010b}%
  \BibitemOpen
  \bibfield  {author} {\bibinfo {author} {\bibfnamefont {Max~A.}\ \bibnamefont
  {Metlitski}}\ and\ \bibinfo {author} {\bibfnamefont {Subir}\ \bibnamefont
  {Sachdev}},\ }\bibfield  {title} {\enquote {\bibinfo {title} {Quantum phase
  transitions of metals in two spatial dimensions. ii. spin density wave
  order},}\ }\href {\doibase 10.1103/PhysRevB.82.075128} {\bibfield  {journal}
  {\bibinfo  {journal} {Phys. Rev. B}\ }\textbf {\bibinfo {volume} {82}},\
  \bibinfo {pages} {075128} (\bibinfo {year} {2010}{\natexlab{b}})}\BibitemShut
  {NoStop}%
\bibitem [{\citenamefont {Liu}\ \emph {et~al.}(2018)\citenamefont {Liu},
  \citenamefont {Xu}, \citenamefont {Qi}, \citenamefont {Sun},\ and\
  \citenamefont {Meng}}]{PhysRevB.98.045116}%
  \BibitemOpen
  \bibfield  {author} {\bibinfo {author} {\bibfnamefont {Zi~Hong}\ \bibnamefont
  {Liu}}, \bibinfo {author} {\bibfnamefont {Xiao~Yan}\ \bibnamefont {Xu}},
  \bibinfo {author} {\bibfnamefont {Yang}\ \bibnamefont {Qi}}, \bibinfo
  {author} {\bibfnamefont {Kai}\ \bibnamefont {Sun}}, \ and\ \bibinfo {author}
  {\bibfnamefont {Zi~Yang}\ \bibnamefont {Meng}},\ }\bibfield  {title}
  {\enquote {\bibinfo {title} {Itinerant quantum critical point with
  frustration and a non-fermi liquid},}\ }\href {\doibase
  10.1103/PhysRevB.98.045116} {\bibfield  {journal} {\bibinfo  {journal} {Phys.
  Rev. B}\ }\textbf {\bibinfo {volume} {98}},\ \bibinfo {pages} {045116}
  (\bibinfo {year} {2018})}\BibitemShut {NoStop}%
\bibitem [{\citenamefont {Liu}\ \emph {et~al.}(2019)\citenamefont {Liu},
  \citenamefont {Pan}, \citenamefont {Xu}, \citenamefont {Sun},\ and\
  \citenamefont {Meng}}]{Liu2019PNAS}%
  \BibitemOpen
  \bibfield  {author} {\bibinfo {author} {\bibfnamefont {Zi~Hong}\ \bibnamefont
  {Liu}}, \bibinfo {author} {\bibfnamefont {Gaopei}\ \bibnamefont {Pan}},
  \bibinfo {author} {\bibfnamefont {Xiao~Yan}\ \bibnamefont {Xu}}, \bibinfo
  {author} {\bibfnamefont {Kai}\ \bibnamefont {Sun}}, \ and\ \bibinfo {author}
  {\bibfnamefont {Zi~Yang}\ \bibnamefont {Meng}},\ }\bibfield  {title}
  {\enquote {\bibinfo {title} {Itinerant quantum critical point with fermion
  pockets and hotspots},}\ }\href {\doibase 10.1073/pnas.1901751116} {\bibfield
   {journal} {\bibinfo  {journal} {Proceedings of the National Academy of
  Sciences}\ }\textbf {\bibinfo {volume} {116}},\ \bibinfo {pages}
  {16760--16767} (\bibinfo {year} {2019})}\BibitemShut {NoStop}%
\bibitem [{\citenamefont {Xu}\ \emph {et~al.}(2019)\citenamefont {Xu},
  \citenamefont {Liu}, \citenamefont {Pan}, \citenamefont {Qi}, \citenamefont
  {Sun},\ and\ \citenamefont {Meng}}]{XiaoYanXuReview2019}%
  \BibitemOpen
  \bibfield  {author} {\bibinfo {author} {\bibfnamefont {Xiao~Yan}\
  \bibnamefont {Xu}}, \bibinfo {author} {\bibfnamefont {Zi~Hong}\ \bibnamefont
  {Liu}}, \bibinfo {author} {\bibfnamefont {Gaopei}\ \bibnamefont {Pan}},
  \bibinfo {author} {\bibfnamefont {Yang}\ \bibnamefont {Qi}}, \bibinfo
  {author} {\bibfnamefont {Kai}\ \bibnamefont {Sun}}, \ and\ \bibinfo {author}
  {\bibfnamefont {Zi~Yang}\ \bibnamefont {Meng}},\ }\bibfield  {title}
  {\enquote {\bibinfo {title} {Revealing fermionic quantum criticality from new
  monte carlo techniques},}\ }\href {\doibase 10.1088/1361-648x/ab3295}
  {\bibfield  {journal} {\bibinfo  {journal} {Journal of Physics: Condensed
  Matter}\ }\textbf {\bibinfo {volume} {31}},\ \bibinfo {pages} {463001}
  (\bibinfo {year} {2019})}\BibitemShut {NoStop}%
\bibitem [{\citenamefont {{Xu}}\ \emph {et~al.}(2020)\citenamefont {{Xu}},
  \citenamefont {{Klein}}, \citenamefont {{Sun}}, \citenamefont {{Chubukov}},\
  and\ \citenamefont {{Meng}}}]{XiaoYanXu2020}%
  \BibitemOpen
  \bibfield  {author} {\bibinfo {author} {\bibfnamefont {Xiao~Yan}\
  \bibnamefont {{Xu}}}, \bibinfo {author} {\bibfnamefont {Avraham}\
  \bibnamefont {{Klein}}}, \bibinfo {author} {\bibfnamefont {Kai}\ \bibnamefont
  {{Sun}}}, \bibinfo {author} {\bibfnamefont {Andrey~V.}\ \bibnamefont
  {{Chubukov}}}, \ and\ \bibinfo {author} {\bibfnamefont {Zi~Yang}\
  \bibnamefont {{Meng}}},\ }\bibfield  {title} {\enquote {\bibinfo {title}
  {{Identification of non-Fermi liquid fermionic self-energy from quantum Monte
  Carlo data}},}\ }\href {\doibase 10.1038/s41535-020-00266-6} {\bibfield
  {journal} {\bibinfo  {journal} {npj Quantum Materials}\ }\textbf {\bibinfo
  {volume} {5}},\ \bibinfo {pages} {65} (\bibinfo {year} {2020})}\BibitemShut
  {NoStop}%
\bibitem [{\citenamefont {Damia}\ \emph {et~al.}(2020)\citenamefont {Damia},
  \citenamefont {Sol\'{\i}s},\ and\ \citenamefont {Torroba}}]{Damia2020}%
  \BibitemOpen
  \bibfield  {author} {\bibinfo {author} {\bibfnamefont {Jeremias~Aguilera}\
  \bibnamefont {Damia}}, \bibinfo {author} {\bibfnamefont {Mario}\ \bibnamefont
  {Sol\'{\i}s}}, \ and\ \bibinfo {author} {\bibfnamefont {Gonzalo}\
  \bibnamefont {Torroba}},\ }\bibfield  {title} {\enquote {\bibinfo {title}
  {How non-fermi liquids cure their infrared divergences},}\ }\href {\doibase
  10.1103/PhysRevB.102.045147} {\bibfield  {journal} {\bibinfo  {journal}
  {Phys. Rev. B}\ }\textbf {\bibinfo {volume} {102}},\ \bibinfo {pages}
  {045147} (\bibinfo {year} {2020})}\BibitemShut {NoStop}%
\bibitem [{\citenamefont {Guo}\ \emph {et~al.}(2020)\citenamefont {Guo},
  \citenamefont {Gu},\ and\ \citenamefont {Sachdev}}]{guo-gu-sachdev-2020}%
  \BibitemOpen
  \bibfield  {author} {\bibinfo {author} {\bibfnamefont {Haoyu}\ \bibnamefont
  {Guo}}, \bibinfo {author} {\bibfnamefont {Yingfei}\ \bibnamefont {Gu}}, \
  and\ \bibinfo {author} {\bibfnamefont {Subir}\ \bibnamefont {Sachdev}},\
  }\bibfield  {title} {\enquote {\bibinfo {title} {Linear in temperature
  resistivity in the limit of zero temperature from the time reparameterization
  soft mode},}\ }\href@noop {} {\bibfield  {journal} {\bibinfo  {journal}
  {Annals of Physics}\ ,\ \bibinfo {pages} {168202}} (\bibinfo {year}
  {2020})}\BibitemShut {NoStop}%
\bibitem [{\citenamefont {Gu}\ \emph {et~al.}(2017{\natexlab{b}})\citenamefont
  {Gu}, \citenamefont {Qi},\ and\ \citenamefont {Stanford}}]{YingfeiGu2017}%
  \BibitemOpen
  \bibfield  {author} {\bibinfo {author} {\bibfnamefont {Yingfei}\ \bibnamefont
  {Gu}}, \bibinfo {author} {\bibfnamefont {Xiao-Liang}\ \bibnamefont {Qi}}, \
  and\ \bibinfo {author} {\bibfnamefont {Douglas}\ \bibnamefont {Stanford}},\
  }\bibfield  {title} {\enquote {\bibinfo {title} {Local criticality, diffusion
  and chaos in generalized {Sachdev-Ye-Kitaev} models},}\ }\href {\doibase
  10.1007/JHEP05(2017)125} {\bibfield  {journal} {\bibinfo  {journal} {Journal
  of High Energy Physics}\ }\textbf {\bibinfo {volume} {2017}},\ \bibinfo
  {pages} {125} (\bibinfo {year} {2017}{\natexlab{b}})}\BibitemShut {NoStop}%
\bibitem [{\citenamefont {Wang}(2020)}]{PhysRevLett.124.017002}%
  \BibitemOpen
  \bibfield  {author} {\bibinfo {author} {\bibfnamefont {Yuxuan}\ \bibnamefont
  {Wang}},\ }\bibfield  {title} {\enquote {\bibinfo {title} {Solvable
  strong-coupling quantum-dot model with a non-fermi-liquid pairing
  transition},}\ }\href {\doibase 10.1103/PhysRevLett.124.017002} {\bibfield
  {journal} {\bibinfo  {journal} {Phys. Rev. Lett.}\ }\textbf {\bibinfo
  {volume} {124}},\ \bibinfo {pages} {017002} (\bibinfo {year}
  {2020})}\BibitemShut {NoStop}%
\bibitem [{\citenamefont {Esterlis}\ and\ \citenamefont
  {Schmalian}(2019)}]{PhysRevB.100.115132}%
  \BibitemOpen
  \bibfield  {author} {\bibinfo {author} {\bibfnamefont {Ilya}\ \bibnamefont
  {Esterlis}}\ and\ \bibinfo {author} {\bibfnamefont {J\"org}\ \bibnamefont
  {Schmalian}},\ }\bibfield  {title} {\enquote {\bibinfo {title} {Cooper
  pairing of incoherent electrons: An electron-phonon version of the
  sachdev-ye-kitaev model},}\ }\href {\doibase 10.1103/PhysRevB.100.115132}
  {\bibfield  {journal} {\bibinfo  {journal} {Phys. Rev. B}\ }\textbf {\bibinfo
  {volume} {100}},\ \bibinfo {pages} {115132} (\bibinfo {year}
  {2019})}\BibitemShut {NoStop}%
\bibitem [{\citenamefont {Hauck}\ \emph {et~al.}(2020)\citenamefont {Hauck},
  \citenamefont {Klug}, \citenamefont {Esterlis},\ and\ \citenamefont
  {Schmalian}}]{HAUCK2020168120}%
  \BibitemOpen
  \bibfield  {author} {\bibinfo {author} {\bibfnamefont {Daniel}\ \bibnamefont
  {Hauck}}, \bibinfo {author} {\bibfnamefont {Markus~J.}\ \bibnamefont {Klug}},
  \bibinfo {author} {\bibfnamefont {Ilya}\ \bibnamefont {Esterlis}}, \ and\
  \bibinfo {author} {\bibfnamefont {Jörg}\ \bibnamefont {Schmalian}},\
  }\bibfield  {title} {\enquote {\bibinfo {title} {Eliashberg equations for an
  electron–phonon version of the sachdev–ye–kitaev model: Pair breaking
  in non-fermi liquid superconductors},}\ }\href {\doibase
  https://doi.org/10.1016/j.aop.2020.168120} {\bibfield  {journal} {\bibinfo
  {journal} {Annals of Physics}\ }\textbf {\bibinfo {volume} {417}},\ \bibinfo
  {pages} {168120} (\bibinfo {year} {2020})},\ \bibinfo {note} {eliashberg
  theory at 60: Strong-coupling superconductivity and beyond}\BibitemShut
  {NoStop}%
\bibitem [{\citenamefont {Pan}\ \emph {et~al.}(2021)\citenamefont {Pan},
  \citenamefont {Wang}, \citenamefont {Davis}, \citenamefont {Wang},\ and\
  \citenamefont {Meng}}]{2020Yukawa}%
  \BibitemOpen
  \bibfield  {author} {\bibinfo {author} {\bibfnamefont {Gaopei}\ \bibnamefont
  {Pan}}, \bibinfo {author} {\bibfnamefont {Wei}\ \bibnamefont {Wang}},
  \bibinfo {author} {\bibfnamefont {Andrew}\ \bibnamefont {Davis}}, \bibinfo
  {author} {\bibfnamefont {Yuxuan}\ \bibnamefont {Wang}}, \ and\ \bibinfo
  {author} {\bibfnamefont {Zi~Yang}\ \bibnamefont {Meng}},\ }\bibfield  {title}
  {\enquote {\bibinfo {title} {{Yukawa-SYK model and self-tuned quantum
  criticality}},}\ }\href {\doibase 10.1103/PhysRevResearch.3.013250}
  {\bibfield  {journal} {\bibinfo  {journal} {Phys. Rev. Research}\ }\textbf
  {\bibinfo {volume} {3}},\ \bibinfo {pages} {013250} (\bibinfo {year}
  {2021})}\BibitemShut {NoStop}%
\bibitem [{\citenamefont {{Kim}}\ \emph
  {et~al.}(2020{\natexlab{a}})\citenamefont {{Kim}}, \citenamefont {{Cao}},\
  and\ \citenamefont {{Altman}}}]{Kim-Cao-Altman}%
  \BibitemOpen
  \bibfield  {author} {\bibinfo {author} {\bibfnamefont {Jaewon}\ \bibnamefont
  {{Kim}}}, \bibinfo {author} {\bibfnamefont {Xiangyu}\ \bibnamefont {{Cao}}},
  \ and\ \bibinfo {author} {\bibfnamefont {Ehud}\ \bibnamefont {{Altman}}},\
  }\bibfield  {title} {\enquote {\bibinfo {title} {{Low-rank Sachdev-Ye-Kitaev
  models}},}\ }\href {\doibase 10.1103/PhysRevB.101.125112} {\bibfield
  {journal} {\bibinfo  {journal} {\prb}\ }\textbf {\bibinfo {volume} {101}},\
  \bibinfo {eid} {125112} (\bibinfo {year} {2020}{\natexlab{a}})},\ \Eprint
  {http://arxiv.org/abs/1910.10173} {arXiv:1910.10173 [cond-mat.str-el]}
  \BibitemShut {NoStop}%
\bibitem [{\citenamefont {{Kim}}\ \emph
  {et~al.}(2020{\natexlab{b}})\citenamefont {{Kim}}, \citenamefont {{Altman}},\
  and\ \citenamefont {{Cao}}}]{Kim-Alterman-Cao}%
  \BibitemOpen
  \bibfield  {author} {\bibinfo {author} {\bibfnamefont {Jaewon}\ \bibnamefont
  {{Kim}}}, \bibinfo {author} {\bibfnamefont {Ehud}\ \bibnamefont {{Altman}}},
  \ and\ \bibinfo {author} {\bibfnamefont {Xiangyu}\ \bibnamefont {{Cao}}},\
  }\bibfield  {title} {\enquote {\bibinfo {title} {{Dirac Fast Scramblers}},}\
  }\href@noop {} {\bibfield  {journal} {\bibinfo  {journal} {arXiv e-prints}\
  ,\ \bibinfo {eid} {arXiv:2010.10545}} (\bibinfo {year}
  {2020}{\natexlab{b}})},\ \Eprint {http://arxiv.org/abs/2010.10545}
  {arXiv:2010.10545 [cond-mat.str-el]} \BibitemShut {NoStop}%
\bibitem [{\citenamefont {Azeyanagi}\ \emph {et~al.}(2018)\citenamefont
  {Azeyanagi}, \citenamefont {Ferrari},\ and\ \citenamefont
  {Massolo}}]{PhysRevLett.120.061602}%
  \BibitemOpen
  \bibfield  {author} {\bibinfo {author} {\bibfnamefont {Tatsuo}\ \bibnamefont
  {Azeyanagi}}, \bibinfo {author} {\bibfnamefont {Frank}\ \bibnamefont
  {Ferrari}}, \ and\ \bibinfo {author} {\bibfnamefont {Fidel I.~Schaposnik}\
  \bibnamefont {Massolo}},\ }\bibfield  {title} {\enquote {\bibinfo {title}
  {Phase diagram of planar matrix quantum mechanics, tensor, and
  sachdev-ye-kitaev models},}\ }\href {\doibase 10.1103/PhysRevLett.120.061602}
  {\bibfield  {journal} {\bibinfo  {journal} {Phys. Rev. Lett.}\ }\textbf
  {\bibinfo {volume} {120}},\ \bibinfo {pages} {061602} (\bibinfo {year}
  {2018})}\BibitemShut {NoStop}%
\bibitem [{\citenamefont {Smit}\ \emph {et~al.}(2020)\citenamefont {Smit},
  \citenamefont {Valentinis}, \citenamefont {Schmalian},\ and\ \citenamefont
  {Kopietz}}]{smit2020quantum}%
  \BibitemOpen
  \bibfield  {author} {\bibinfo {author} {\bibfnamefont {Roman}\ \bibnamefont
  {Smit}}, \bibinfo {author} {\bibfnamefont {Davide}\ \bibnamefont
  {Valentinis}}, \bibinfo {author} {\bibfnamefont {J{\"o}rg}\ \bibnamefont
  {Schmalian}}, \ and\ \bibinfo {author} {\bibfnamefont {Peter}\ \bibnamefont
  {Kopietz}},\ }\bibfield  {title} {\enquote {\bibinfo {title} {Quantum
  discontinuity fixed point and renormalization group flow of the syk model},}\
  }\href@noop {} {\bibfield  {journal} {\bibinfo  {journal} {arXiv preprint
  arXiv:2010.01142}\ } (\bibinfo {year} {2020})}\BibitemShut {NoStop}%
\bibitem [{\citenamefont {Wang}\ \emph {et~al.}(2020)\citenamefont {Wang},
  \citenamefont {Chudnovskiy}, \citenamefont {Gorsky},\ and\ \citenamefont
  {Kamenev}}]{PhysRevResearch.2.033025}%
  \BibitemOpen
  \bibfield  {author} {\bibinfo {author} {\bibfnamefont {Hanteng}\ \bibnamefont
  {Wang}}, \bibinfo {author} {\bibfnamefont {A.~L.}\ \bibnamefont
  {Chudnovskiy}}, \bibinfo {author} {\bibfnamefont {Alexander}\ \bibnamefont
  {Gorsky}}, \ and\ \bibinfo {author} {\bibfnamefont {Alex}\ \bibnamefont
  {Kamenev}},\ }\bibfield  {title} {\enquote {\bibinfo {title}
  {Sachdev-ye-kitaev superconductivity: Quantum kuramoto and generalized
  richardson models},}\ }\href {\doibase 10.1103/PhysRevResearch.2.033025}
  {\bibfield  {journal} {\bibinfo  {journal} {Phys. Rev. Research}\ }\textbf
  {\bibinfo {volume} {2}},\ \bibinfo {pages} {033025} (\bibinfo {year}
  {2020})}\BibitemShut {NoStop}%
\bibitem [{\citenamefont {Setty}(2020)}]{PhysRevB.101.184506}%
  \BibitemOpen
  \bibfield  {author} {\bibinfo {author} {\bibfnamefont {Chandan}\ \bibnamefont
  {Setty}},\ }\bibfield  {title} {\enquote {\bibinfo {title} {Pairing
  instability on a luttinger surface: A non-fermi liquid to superconductor
  transition and its sachdev-ye-kitaev dual},}\ }\href {\doibase
  10.1103/PhysRevB.101.184506} {\bibfield  {journal} {\bibinfo  {journal}
  {Phys. Rev. B}\ }\textbf {\bibinfo {volume} {101}},\ \bibinfo {pages}
  {184506} (\bibinfo {year} {2020})}\BibitemShut {NoStop}%
\bibitem [{\citenamefont {Cheipesh}\ \emph {et~al.}(2019)\citenamefont
  {Cheipesh}, \citenamefont {Pavlov}, \citenamefont {Scopelliti}, \citenamefont
  {Tworzyd\l{}o},\ and\ \citenamefont {Gnezdilov}}]{PhysRevB.100.220506}%
  \BibitemOpen
  \bibfield  {author} {\bibinfo {author} {\bibfnamefont {Y.}~\bibnamefont
  {Cheipesh}}, \bibinfo {author} {\bibfnamefont {A.~I.}\ \bibnamefont
  {Pavlov}}, \bibinfo {author} {\bibfnamefont {V.}~\bibnamefont {Scopelliti}},
  \bibinfo {author} {\bibfnamefont {J.}~\bibnamefont {Tworzyd\l{}o}}, \ and\
  \bibinfo {author} {\bibfnamefont {N.~V.}\ \bibnamefont {Gnezdilov}},\
  }\bibfield  {title} {\enquote {\bibinfo {title} {Reentrant superconductivity
  in a quantum dot coupled to a sachdev-ye-kitaev metal},}\ }\href {\doibase
  10.1103/PhysRevB.100.220506} {\bibfield  {journal} {\bibinfo  {journal}
  {Phys. Rev. B}\ }\textbf {\bibinfo {volume} {100}},\ \bibinfo {pages}
  {220506} (\bibinfo {year} {2019})}\BibitemShut {NoStop}%
\bibitem [{\citenamefont {Imada}\ \emph {et~al.}(1998)\citenamefont {Imada},
  \citenamefont {Fujimori},\ and\ \citenamefont {Tokura}}]{Imada1998}%
  \BibitemOpen
  \bibfield  {author} {\bibinfo {author} {\bibfnamefont {Masatoshi}\
  \bibnamefont {Imada}}, \bibinfo {author} {\bibfnamefont {Atsushi}\
  \bibnamefont {Fujimori}}, \ and\ \bibinfo {author} {\bibfnamefont
  {Yoshinori}\ \bibnamefont {Tokura}},\ }\bibfield  {title} {\enquote {\bibinfo
  {title} {Metal-insulator transitions},}\ }\href {\doibase
  10.1103/RevModPhys.70.1039} {\bibfield  {journal} {\bibinfo  {journal} {Rev.
  Mod. Phys.}\ }\textbf {\bibinfo {volume} {70}},\ \bibinfo {pages}
  {1039--1263} (\bibinfo {year} {1998})}\BibitemShut {NoStop}%
\bibitem [{\citenamefont {Limelette}\ \emph {et~al.}(2003)\citenamefont
  {Limelette}, \citenamefont {Georges}, \citenamefont {J{\'e}rome},
  \citenamefont {Wzietek}, \citenamefont {Metcalf},\ and\ \citenamefont
  {Honig}}]{Limelette2003}%
  \BibitemOpen
  \bibfield  {author} {\bibinfo {author} {\bibfnamefont {P.}~\bibnamefont
  {Limelette}}, \bibinfo {author} {\bibfnamefont {A.}~\bibnamefont {Georges}},
  \bibinfo {author} {\bibfnamefont {D.}~\bibnamefont {J{\'e}rome}}, \bibinfo
  {author} {\bibfnamefont {P.}~\bibnamefont {Wzietek}}, \bibinfo {author}
  {\bibfnamefont {P.}~\bibnamefont {Metcalf}}, \ and\ \bibinfo {author}
  {\bibfnamefont {J.~M.}\ \bibnamefont {Honig}},\ }\bibfield  {title} {\enquote
  {\bibinfo {title} {Universality and critical behavior at the mott
  transition},}\ }\href {\doibase 10.1126/science.1088386} {\bibfield
  {journal} {\bibinfo  {journal} {Science}\ }\textbf {\bibinfo {volume}
  {302}},\ \bibinfo {pages} {89--92} (\bibinfo {year} {2003})}\BibitemShut
  {NoStop}%
\bibitem [{\citenamefont {Isakov}\ and\ \citenamefont
  {Moessner}(2003)}]{isakov2003interplay}%
  \BibitemOpen
  \bibfield  {author} {\bibinfo {author} {\bibfnamefont {S.~V.}\ \bibnamefont
  {Isakov}}\ and\ \bibinfo {author} {\bibfnamefont {R.}~\bibnamefont
  {Moessner}},\ }\bibfield  {title} {\enquote {\bibinfo {title} {Interplay of
  quantum and thermal fluctuations in a frustrated magnet},}\ }\href {\doibase
  10.1103/PhysRevB.68.104409} {\bibfield  {journal} {\bibinfo  {journal} {Phys.
  Rev. B}\ }\textbf {\bibinfo {volume} {68}},\ \bibinfo {pages} {104409}
  (\bibinfo {year} {2003})}\BibitemShut {NoStop}%
\bibitem [{\citenamefont {Paiva}\ \emph {et~al.}(2004)\citenamefont {Paiva},
  \citenamefont {dos Santos}, \citenamefont {Scalettar},\ and\ \citenamefont
  {Denteneer}}]{Paiva2004}%
  \BibitemOpen
  \bibfield  {author} {\bibinfo {author} {\bibfnamefont {Thereza}\ \bibnamefont
  {Paiva}}, \bibinfo {author} {\bibfnamefont {Raimundo~R.}\ \bibnamefont {dos
  Santos}}, \bibinfo {author} {\bibfnamefont {R.~T.}\ \bibnamefont
  {Scalettar}}, \ and\ \bibinfo {author} {\bibfnamefont {P.~J.~H.}\
  \bibnamefont {Denteneer}},\ }\bibfield  {title} {\enquote {\bibinfo {title}
  {Critical temperature for the two-dimensional attractive hubbard model},}\
  }\href {\doibase 10.1103/PhysRevB.69.184501} {\bibfield  {journal} {\bibinfo
  {journal} {Phys. Rev. B}\ }\textbf {\bibinfo {volume} {69}},\ \bibinfo
  {pages} {184501} (\bibinfo {year} {2004})}\BibitemShut {NoStop}%
\bibitem [{\citenamefont {Costa}\ \emph {et~al.}(2018)\citenamefont {Costa},
  \citenamefont {Blommel}, \citenamefont {Chiu}, \citenamefont {Batrouni},\
  and\ \citenamefont {Scalettar}}]{costa2018phonon}%
  \BibitemOpen
  \bibfield  {author} {\bibinfo {author} {\bibfnamefont {N.~C.}\ \bibnamefont
  {Costa}}, \bibinfo {author} {\bibfnamefont {T.}~\bibnamefont {Blommel}},
  \bibinfo {author} {\bibfnamefont {W.-T.}\ \bibnamefont {Chiu}}, \bibinfo
  {author} {\bibfnamefont {G.}~\bibnamefont {Batrouni}}, \ and\ \bibinfo
  {author} {\bibfnamefont {R.~T.}\ \bibnamefont {Scalettar}},\ }\bibfield
  {title} {\enquote {\bibinfo {title} {Phonon dispersion and the competition
  between pairing and charge order},}\ }\href {\doibase
  10.1103/PhysRevLett.120.187003} {\bibfield  {journal} {\bibinfo  {journal}
  {Phys. Rev. Lett.}\ }\textbf {\bibinfo {volume} {120}},\ \bibinfo {pages}
  {187003} (\bibinfo {year} {2018})}\BibitemShut {NoStop}%
\bibitem [{Note1()}]{Note1}%
  \BibitemOpen
  \bibinfo {note} {For a mean field theory in finite spatial dimensions,
  Josephson's identity states that $\nu d =2$, and in our model the
  dimensionality $d$ does not enter the theory. We have instead $\nu =2$. We
  thank Ilya Esterlis and Joerg Schmalian for sharing their unpublished results
  with us on this.}\BibitemShut {Stop}%
\end{thebibliography}%


\end{document}